\documentclass[preprint]{aastex}
\slugcomment{submitted to ApJ}

\shorttitle{Chemical evolution along the Hubble Sequence}
\shortauthors{Moll\´{a} et al.}
\input epsf
\usepackage{graphicx}

\begin{document}

\title{Galactic evolution along the Hubble Sequence\\
I. A grid of models parametrized by initial galaxy mass distribution}

\author{Mercedes Moll\'{a}, Angeles I. D\'{\i}az}
\affil{Departamento de F\'{\i}sica Te\'{o}rica,
Universidad Aut\'onoma de Madrid, 28049 Cantoblanco}
\email{mercedes.molla@uam.es}
\email{angeles.diaz@uam.es}

\author{Federico Ferrini\altaffilmark{1}}
\affil{Dipartimento di Fisica,
Universit\`{a} di Pisa, Piazza Torricelli 2, 56100 Pisa, Italia }
\altaffiltext{1}{INTAS, 58, Avenue des Arts, 1000 Bruxelles, Belgium} 
\email{ferrini@intas.be}

\begin{abstract}

We present a generalization of the multiphase chemical evolution model
applied to a wide set of theoretical galaxies with different masses
and morphological types.  This generalized set of models has been
computed using the so-called {\sl Universal Rotation Curve} from
\citet{per96} to calculate the radial mass distribution of 44 {\sl
theoretical} protogalaxies. This distribution is a fundamental input
which, besides its own effect on the galaxy evolution, defines the
characteristic collapse time scale or gas infall rate onto the disk.

 On the other hand, the molecular cloud and star formation
efficiencies take 10 different values between 0 and 1, as
corresponding to their probability nature, for each one of the radial
distributions of total mass. This implies that for each mass radial
distribution, we have 10 different evolutionary rates, which happen to
be related to the morphological types of galaxies, as we will later
show. With these two hypotheses we construct a bi-parametric grid of
models whose results are valid in principle for any spiral galaxy of
given maximum rotation velocity or total mass, and type T from 1 to
10.

The results include the time evolution of different regions of the
disk and the halo along galactocentric distance, measured by the gas
(atomic and molecular) and stellar masses, the star formation rate and
chemical abundances of 15 elements. The present time radial
distributions of diffuse and molecular gas and star formation rate
surface densities, and of oxygen abundance - defined as 12 +
log(O/H)-- are calculated and then compared with observational data.

One of the most important results of this work concerns the radial
gradients of abundances. These are completely flat for the latest
morphological types, from $T\sim$ 7 to 10, with abundances $12+ log
(O/H) \sim 7.5-8$, while they appear steep for late type galaxies (T
$\sim 4-7$), in comparison when earlier types. This is in agreement
with observations and it resolves the apparent inconsistency in the
trends giving larger gradients for later type galaxies while some
irregulars show no gradient at all and very uniform abundances. The
explanation resides in the star formation to infall ratio provided by
the multiphase models.

These models are also able to explain the existing correlations
between generic characteristics of galaxies and their radial
distributions, as arising from variations in the characteristic infall
rate and in the cloud and star formation efficiencies with galactic
morphological type and/or Arm Class.

\end{abstract}


\section{Introduction}

Early chemical evolution models \citep{lyn75,tin80,cla87,cla88,som89}
were developed with the aim of explaining the solar elemental
abundances, now revised in \citet{gre98}, and other characteristics
observed in the Solar Neighborhood, such as the metallicity
distribution \citep{pag75,pag90,rocha96,chang99}, the age-metallicity
relation of stars or the star formation history
\citep{twa80,car85,bar88,rocha00,rocha00b}.  Subsequently, the scope
of the models was enlarged to include the whole Galactic Disk, in an
attempt to reproduce the observed radial gradients in the abundances
of oxygen, nitrogen and other elements, as deduced from H{\sc ii}
regions and planetary nebulae (PN) observations \citep{pei79,sha83}
and of iron, as derived from data on stars and open clusters.

Variations of abundances with galactocentric distance have also been
found, mainly from observations of H\,{\sc ii} regions, in other
spirals \citep[see][and references therein]{hen99}, and the existence
of abundance radial gradients in spiral disks has been firmly
established.

From these data, several correlations have been found between the
values of these gradients and galaxy characteristics, {\sl e.g.}
\citet{vil92,zar94}, where a first quantification of these trends was
done.  We can, thus, enumerate the following observational facts:

\begin{enumerate}
\item The central value of the abundance, as extrapolated from the
obtained gradient, depends on the morphological type with the earlier
type galaxies showing larger central abundances than the later ones
\citep{vil92,zar94}.

\item The radial gradient in dex $\rm kpc^{-1}$ depends on the
morphological type: late-type galaxies show steeper radial
distributions of oxygen abundances than earlier ones, which show in
some cases almost no gradient  \citep{oey93,dut99}.

\item There is a correlation between abundance and total mass surface
density, which, in turn, is related to the morphological type. This
correlation is stronger when the total mass of the galaxy, including
the bulge, instead of the mass of the exponential disk alone, is used
\citep{ryd95}.

\item A correlation exists between the atomic gas fraction and the
oxygen abundance, which remains when the molecular gas component is
also included, but only if this mass is estimated using a conversion
factor $\chi$ of CO intensity to molecular gas density variable with
metallicity in a way that produces a smooth radial distribution of gas
\citep{vil92}.

\item The fraction H$_2$/HI varies with the oxygen abundance, as it
was suggested by \citet{tos90}.
\end{enumerate}

From the point of view of theoretical models, the existence of these
radial gradients can not be due only to the gas fraction radial
distribution, and other effects must be invoked
\citep{tin80,got92,kop94}. Thus, most recently computed models, including
the multiphase model used in this work, assume
a SFR law depending on galactocentric radius through a power law of
the gas density.  They also include the infall of gas whose rate
varies with galactocentric radius. Therefore they follow the so-called
{\sl biased infall} hypothesis, where the disk forms in an
inside/outside scenario. The physical reason for this is that the gas
tends to collapse more quickly in the inner regions of the disk, the
process being slower in the outer regions.  These models are very
similar in their hypotheses and results for the present time.

Most numerical models, although more sophisticated every time, have
been historically checked only on the MWG or a fiducial galaxy as an
example. There have been models computed to analyze irregular galaxies
such as the Magellanic Clouds \citep{pag98} or IZw18. However, with
the exception of \citet{dia84} and \citet{tos85}, who modeled a sample
of external spiral galaxies using rather simple assumptions about the
star formation and infall rates, chemical evolution models have not
been usually applied to other external large spirals.  This prevents
the adequate comparison of the correlations mentioned above with model
results. This is the main purpose of this work.

The multiphase model we will use here, has been already used and
checked against observational constraints not only for the Milky Way
Galaxy, as it is commonly done, but also for a sample of spiral
galaxies (disks and bulges) \citep{mol96,mol99,mol00} of different
morphological types and total masses, and the observed radial
distributions of diffuse and molecular gas, oxygen abundances and star
formation rate have been reproduced rather successfully.

The characteristics of the multiphase model have been described in
\citet{fer92}, \citet[][--hereafter FMPD--]{fer94}, and \citet{mol95}.
It follows the two hypotheses above mentioned (SF and infall depending
on the radius), as assumed in other works \citep{boi00,hou00}.

An advantage of using our model is that it includes a more realistic
star formation. The usual SFR prescriptions are based in a Schmidt
law, depending on the total gas surface density. Instead, the
multiphase model assumes a star formation which takes place in
two-steps: first, the formation of molecular clouds; then the
formation of stars.  This simulates a power law on the gas density
with an exponent $n> 1$ with a threshold gas density as shown by
\citet{ken89}, and, more important, it allows the calculation of the
two different gas phases present in the interstellar medium. In fact,
the actual process of star formation, born by observations
\citep{kle01}, is closer to our scenario with stars forming in regions
where there are molecular clouds, than to the classical Schmidt law
which depends only on the total gas density. Our assumed SFR implies
that the some feedback mechanisms are included naturally and
are sufficient to simulate the actually observed process of creation
of stars from the interstellar medium. 

To know the chemical evolution along the Hubble Sequence, that is, for
spiral galaxies with different morphological types, we have also
analyzed the observed correlations between the abundance gradients and
galaxy characteristics. They are well reproduced for a reduced sample
of galaxies \citep{mol96,mol99c}, and are interpreted as arising from
variations in the characteristic infall rate and in the cloud and star
formation efficiencies with galactic morphological type and/or Arm
Class.  However, some of these correlations may pose problems when the
end of the Hubble Sequence is included. Thus, the mass-metallicity
relation seems to be the same for bright spirals and low-mass
irregular galaxies, but the correlation of a steep radial gradient of
the oxygen abundance for galaxies of later types shows an effect
on-off: when the galaxy mass decreases or the Hubble type changes to
the latest irregulars, the steep radial gradient disappears and the
abundance pattern becomes uniform for the whole disk \citep{wal97}, a
result difficult to explain.

Besides that, the number of galaxies modeled until now is small (11)
and restricted in morphological type (T varies between 3 and 7) and
luminosity class (mostly bright spirals). This situation is unlikely
to change since due to the method of modeling, by fitting a large
number of constraints for every galaxy, it is not possible to model a
large number of individual objects. 

In order to extend our modeling to the whole parameter space, we have
used the Universal Rotation Curve from \citet{per96} to calculate a
large number of radial mass distributions.  These distributions, which
are the fundamental input of the multiphase model, represent
theoretical protogalaxies or initial structures which would evolve to
form the observed disks or irregulars.  The total mass of each
simulated galaxy, besides its own effect on the galaxy evolution,
defines the characteristic collapse time scale or gas infall rate onto
the disk. On the other hand we assume that the molecular cloud and
star formation efficiencies are probabilities with values between 0
and 1 and thus, we select 10 different set of values. With this two
hypotheses we construct a bi-parametric grid of 440 models simulating
galaxies of 44 different total masses which may evolve, each one of
them, at 10 different rates, which, as we will see, may be related on 
morphological types from T$=1$ to T$=10$.

This grid implies the computation of a large number of models,
and extends our previous ones which were only applied to a reduced
number of galaxies (11). With this new grid we analyze the evolution
of galaxies having any possible combination of total mass and
efficiencies to form stars and molecular clouds. Moreover, the
statistical analysis of model results and their comparison with
correlations may now be performed more adequately. These models could
also be applied to theoretical galaxies resulting from self-consistent
hydrodynamical numerical simulations thus following their subsequent
chemical evolution.

With this series of models we try to study the time evolution
of galaxies of different masses and morphological types. No attempt
has been made, however, to explain how the morphological type sequence
is produced. As we use a chemical evolution code, we cannot obtain any
dynamical information, such as velocity dispersions, bulge sizes, or
spiral pattern characteristics, necessary to determine the
morphological type from our resulting galaxies. But we can, as we do,
compare our results with data of actual galaxies of given
morphological types and check if they are well reproduced. 

In Section 2 we summarize the generic multiphase chemical evolution
model characteristics and the strategy of its application to spirals
of different types. We analyze our results in a general way
in Section 3.  In Section 4 we will show how the model results are in
agreement with the observations for some individual
galaxies. Finally, the conclusions of this work are presented in
Section 5.

Our results, such as it occurs with most numerical calculations, at
difference to analytical methods, have been until now only shown in a
graphical way, thus making difficult their possible use by the
scientific community not directly involved in the chemical evolution
field.  For this purpose all results obtained in this work will be
available in electronic form at CDS via anonymous ftp to
cdsarc.u-strasbg.fr (130.79.128.5) or via 
{http://cdsweb.u-strasbg.fr/Abstract.html}, or
{http://pollux.ft.uam.es/astro/mercedes/grid} or upon request to
authors.

\section{The Generic Multiphase Model}

In the generalization of this model we assume a protogalaxy to be a
spheroid composed of primordial gas whose total mass and radial
distribution M(R) is calculated from its corresponding rotation curve.
The Universal Rotation Curve from \citet[][hereafter PSS]{per96} has
been used for obtaining the required inputs to the models.  These
authors use a homogeneous sample of about 1100 optical and radio
rotation curves to estimate their profile and amplitude which are
analyzed statistically. From this study, they obtain an expression for
the rotation velocity:
  
\begin{equation}
V(R)=V(R_{opt})\left\{ (0.72+0.44 \log{\lambda}) \frac{1.97
x^{1.22}}{(x^{2}+0.61)^{1.43}}+ 1.6 e^{-0.4 \lambda}
\frac{x^{2}}{x^{2}+2.25 \lambda^{0.4}}\right\}^{1/2} \rm km s^{-1}
\end{equation}

where R$_{opt}$ is the radius encompassing 83\% of the total
integrated light, given by the expression: $R_{opt} = 13
(L/L_{*}^{0.5})$ kpc, $x=R/R_{opt}$ and $V(R_{opt})$ is the rotation
velocity at the optical radius $R_{opt}$ in $\rm km s^{-1}$:

\begin{equation}
V(R_{opt})=\frac{200 \lambda^{0.41}}{[0.80 + 0.49 log \lambda + (0.75
e^{-0.4 \lambda})/(0.47 + 2.25 \lambda^{0.4})]^{1/2}} \rm km s^{-1}
\end{equation} 

where the value $\lambda$ represents $L/L_{*}$, with $L_{*} =
10^{10.4}$ L$_{\odot}$.

In Table~\ref{grid_m} we show the characteristics obtained with PSS96
equations for 44 different values of $\lambda$. In Column (1) we give
the number of the radial distribution, defined by the value of
$\lambda$, given in Column (2). The total integrated luminosity $L$
and the corresponding magnitude in I-band are in Columns (3) and (4).
The virial radius $R_{gal}=14.8 \lambda^{-0.14} R_{opt}$, the
exponential disk scale length $R_{D}=R_{opt}/3.2$ and the
characteristic radius, which we will use as our reference radius,
defined for each radial distribution as $R_{0}=R_{opt}/2$ are, in
units of kpc, in Columns (5),(6) and (7), respectively. Column (8)
shows the rotation velocity, in $\rm km s^{-1}$, reached at a radius
$R_{M}=2.2R_{D}$ kpc. From this value we compute
$\Omega_{max}=V_{max}/R_{M}$, given in Column (9), in $\rm km s^{-1}
kpc^{-1}$. The total mass of the galaxy, calculated with the classical
expression $M_{gal}=2.32 10^{5} R_{gal} V_{max}^{2}$, in units of
10$^{9}$ M$_{\odot}$, is in Column (10), and the characteristic
collapse time scale, in Gyr, which will be described below,
corresponding to each distribution, is given in Column (11).

\begin{deluxetable}{rlccccccccc}
\tabletypesize{\scriptsize} \tablecaption{Galaxy Characteristics
dependent on the Total Mass.
\label{grid_m}}     
\tablehead{ \colhead{Num} & \colhead{$\lambda$} & \colhead{log L} &
\colhead{M$_{I}$} & \colhead{R$_{\rm gal}$} & \colhead{R$_{\rm D}$} &
\colhead{R$_{0}$} & \colhead{V$_{\rm max}$}& \colhead{$\Omega_{\rm
max}$}& \colhead{M$_{gal}$} & \colhead{$\tau_{0}$} \\ \colhead{} &
\colhead{} & \colhead{L$_{\odot}$} & \colhead{} & \colhead{kpc} &
\colhead{kpc} & \colhead{kpc} & \colhead{($\rm km s^{-1}$)} &
\colhead{($\rm km s^{-1} kpc^{-1}$)} & \colhead{(10$^{9}M_{\odot}$)} &
\colhead{(Gyr)}} 
\startdata 
1& 0.01& 8.40& -16.53& 36.7& 0.4& 0.7&30.& 34.02& 8.& 60.37\\ 
2& 0.02& 8.70& -17.35& 47.1& 0.6& 0.9& 40.& 31.95& 18.& 40.12\\ 
3& 0.03& 8.88& -17.83& 54.4& 0.7& 1.1& 48.& 30.80& 29.& 31.59\\ 
4& 0.04& 9.00& -18.17& 60.4& 0.8& 1.3& 54.& 30.01& 40.& 26.66\\ 
5& 0.05& 9.10& -18.43& 65.4& 0.9& 1.5& 59.& 29.40& 52.&23.38\\ 
6& 0.06& 9.18& -18.65& 69.9& 1.0& 1.6& 63.& 28.92& 65.&21.00\\ 
7& 0.07& 9.25& -18.83& 73.9& 1.1& 1.7& 67.& 28.52& 78.&19.17\\ 
8& 0.08& 9.30& -18.99& 77.5& 1.1& 1.8& 71.& 28.18& 91.&17.72\\ 
9& 0.09& 9.35& -19.13& 80.9& 1.2& 1.9& 75.& 27.88& 105.&16.53\\ 
10& 0.10& 9.40& -19.25& 84.0& 1.3& 2.1& 78.& 27.61& 119.&15.54\\ 
11& 0.11& 9.44& -19.36& 86.9& 1.3& 2.2& 81.& 27.38& 133.&14.69\\ 
12& 0.12& 9.48& -19.47& 89.7& 1.4& 2.3& 84.& 27.16& 147.&13.96\\ 
13& 0.13& 9.51& -19.56& 92.3& 1.5& 2.3& 87.& 26.96& 162.&13.31\\ 
14& 0.14& 9.55& -19.65& 94.8& 1.5& 2.4& 90.& 26.78& 176.&12.74\\ 
15& 0.15& 9.58& -19.73& 97.2& 1.6& 2.5& 92.& 26.62& 191.&12.24\\ 
16& 0.16& 9.60& -19.80& 99.5& 1.6& 2.6& 95.& 26.46& 207.&11.78\\ 
17& 0.17& 9.63& -19.88& 101.7& 1.7& 2.7& 97.& 26.32& 222.&11.37\\ 
18& 0.18& 9.66& -19.94& 103.8& 1.7& 2.8& 99.& 26.18& 237.&10.99\\ 
19& 0.19& 9.68& -20.01& 105.8& 1.8& 2.8& 101.& 26.05& 253.&10.64\\ 
20& 0.20& 9.70& -20.07& 107.8& 1.8& 2.9& 104.& 25.93& 269.&10.33\\ 
21& 0.30& 9.88& -20.55& 124.7& 2.2& 3.6& 122.& 25.00& 433.&8.13\\ 
22& 0.40& 10.00& -20.89& 138.3& 2.6& 4.1& 138.& 24.35& 608.&6.86\\ 
23& 0.50& 10.10& -21.15& 149.9& 2.9& 4.6& 151.& 23.86& 791.&6.02\\ 
24& 0.60& 10.18& -21.36& 160.1& 3.1& 5.0& 163.& 23.47& 981.&5.41\\ 
25& 0.70& 10.25& -21.55& 169.2& 3.4& 5.4& 173.& 23.15& 1176.&4.94\\ 
26& 0.80& 10.30& -21.70& 177.5& 3.6& 5.8& 183.& 22.87& 1377.&4.56\\ 
27& 0.90& 10.35& -21.84& 185.2& 3.9& 6.2& 192.& 22.63& 1582.&4.26\\ 
28& 1.00& 10.40& -21.97& 192.4& 4.1& 6.5& 200.& 22.41& 1791.&4.00\\ 
29& 1.10& 10.44& -22.08& 199.1& 4.3& 6.8& 208.& 22.22& 2004.&3.78\\ 
30& 1.20& 10.48& -22.18& 205.5& 4.5& 7.1& 216.& 22.04& 2220.&3.59\\ 
31& 1.30& 10.51& -22.28& 211.5& 4.6& 7.4& 223.& 21.88& 2440.&3.43\\ 
32& 1.40& 10.55& -22.36& 217.2& 4.8& 7.7& 230.& 21.74& 2663.&3.28\\ 
33& 1.50& 10.58& -22.45& 222.6& 5.0& 8.0& 236.& 21.60& 2888.&3.15\\ 
34& 1.60& 10.60& -22.52& 227.9& 5.1& 8.2& 243.& 21.48& 3117.&3.03\\ 
35& 1.70& 10.63& -22.59& 232.9& 5.3& 8.5& 249.& 21.36& 3347.&2.93\\ 
36& 1.80& 10.66& -22.66& 237.7& 5.5& 8.7& 255.& 21.25& 3581.&2.83\\ 
37& 1.90& 10.68& -22.72& 242.4& 5.6& 9.0& 260.& 21.15& 3816.&2.74\\ 
38& 2.00& 10.70& -22.79& 246.9& 5.7& 9.2& 266.& 21.05& 4054.&2.66\\ 
39& 2.50& 10.80& -23.05& 267.6& 6.4& 10.3& 291.& 20.63& 5274.&2.33\\ 
40& 3.00& 10.88& -23.26& 285.7& 7.0& 11.3& 314.& 20.29& 6539.&2.09\\ 
41& 3.50& 10.94& -23.45& 302.0& 7.6& 12.2& 335.& 20.01& 7843.&1.91\\ 
42& 4.00& 11.00& -23.60& 316.9& 8.1& 13.0& 353.& 19.77& 9180.&1.77\\ 
43& 4.50& 11.05& -23.74& 330.6& 8.6& 13.8& 371.& 19.56& 10548.&1.65\\ 
44& 5.00& 11.10& -23.87& 343.4& 9.1& 14.5& 387.& 19.37& 11944.&1.55\\ 
\enddata 
\tablecomments{Column 1 is the number of the radial
distribution, defined by the value of $\lambda=L/L_{*}$ ($L_{*}
=10^{10.4}$ L$_{\odot}$), given in Column (2). Columns (3) and (4) are
the total integrated luminosity $L$, in logarithmic scale, and the
corresponding magnitude in I-band.  The virial radius $R_{gal}=14.8
\lambda^{-0.14} R_{opt}$, the exponential disk scale length
$R_{D}=Ropt/3.2$ and the characteristic radius, defined as
$R_{0}=Ropt/2$ are, in units of kpc, in Columns (5),(6) and (7),
respectively. Column (8) shows the rotation velocity, in $\rm km
s^{-1}$, reached at a radius $R_{M}=2.2R_{D}$ kpc.
$\Omega_{max}=V_{max}/R_{M}$, is given in Column (9) in $\rm km s^{-1}
kpc^{-1}$. The total mass of the galaxy $M_{gal}=2.32 10^{5} R_{gal}
V_{max}^{2}$, in units of 10$^{9}$ M$_{\odot}$, is in Column (10), and
the characteristic collapse time scale, in Gyr, is given in Column
(11)}

\end{deluxetable}

\begin{figure}
\plotone{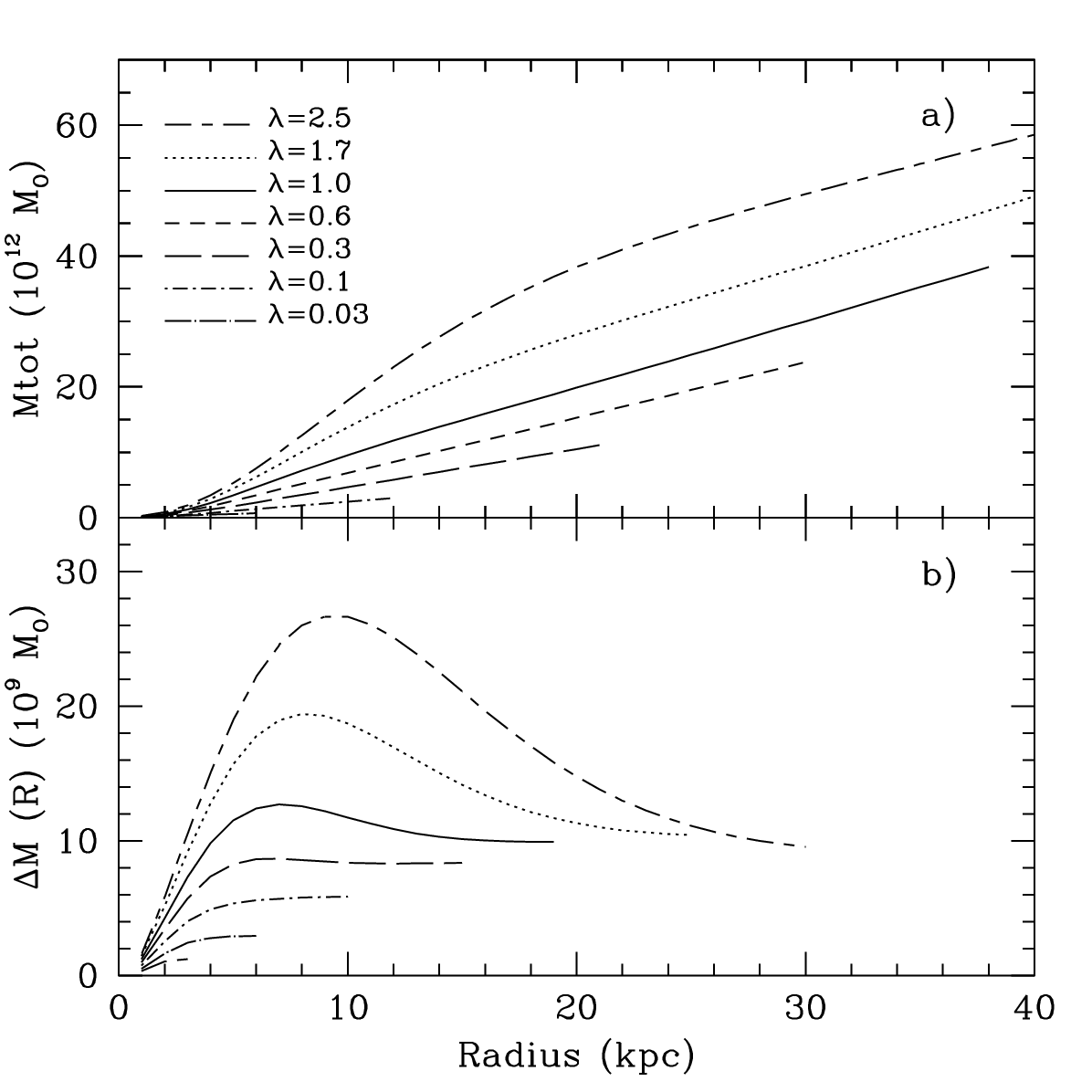}
\caption{Radial distributions: a) total masses M$_{tot}$, b) masses
$\Delta M (R)$ included in our cylinders, for different values of
$\lambda$ following the labels in the figure.}
\label{dis}
\end{figure}

Each galaxy is divided into concentric cylindrical regions 1 kpc wide.
From the corresponding rotation curves we calculate the radial
distributions of total mass M(R), and the total mass included in each
one of cylinders $\Delta M (R)$.  Both distributions are shown in
Fig.\ref{dis}. The total mass includes the dark matter component (DM),
which should not be considered, in principle, in the chemical
evolution calculations. However, maximum disk models have been
suggested by a number of papers \citep[][and references there
in]{pal00,sell00} for decomposition of masses in spiral
galaxies. These works claim that 75\% of the spiral galaxies are well
fitted without a dark matter halo and that the failure to reproduce an
other 20\% is directly related to the existence of non-axysimmetric
structures (bars or strong spiral arms). This implies that the
contribution of DM seems to be negligible in the large massive
galaxies, and more so in the inner parts of disks, where the chemical
evolution takes place.

Once the mass distribution is given, the model calculates, separately,
the time evolution of the halo and disk components belonging to each
cylindrical region. The halo gas falls into the equatorial plane to
form out the disk, which is a secondary structure, created by the
gravitational accumulation of this gas.  The gas infall from the halo
is parameterized by $ f g_{H}$, where $g_{H}$ is the gas mass of the
halo and $f$ is the infall rate, the inverse of the collapse time
scale $\tau$.

The total mass, computed for each rotation curve, defines the
characteristic collapse time scale for each galaxy, through the
expression: $\tau_{0}\propto M_{9}^{-1/2} T_{9}$ \citep{gal84}, where
$M_{9}$ is the total mass of the galaxy in 10$^{9}M_{\odot}$ and
$T_{9}$ is its age, assumed 13 Gyr in all cases.  From the ratio of
the corresponding total mass of a given galaxy and the MWG, and
after calibration with the MWG model, we obtain a value $\tau_{0}$ for
each spiral galaxy:
\begin{equation}
\tau_{0}=\tau_{\odot}(M_{9},gal/M_{9,MWG})^{-1/2}
\label{tau}
\end{equation}

We would like emphasize that the characteristic collapse
time scale computed with Equation~ \ref{tau} is not a free-fall time
but has been calculated through calibration with the Milky Way value,
$\tau_{\odot} \sim 4-6$ Gyr, determined in \citet{fer92} and very
similar to that found in other standard Galactic chemical evolution
models. This value is constrained by using a large number of data,
such as the ratio of the halo to disk mass, $M_{halo}/M_{disk}$, the
relation of [O/Fe] {\sl vs} [Fe/H] for stars in the halo and in the
disk, and the present infall rate, \citep[0.7
M$_{\odot}$pc$^{-2}$Gyr$^{-1}$][]{mir89}, which are well reproduced by
this long scale to form the disk.

The mass which does not fall will remain in the halo, thus
yielding a ratio $M_{halo}/M_{disk}$ for the baryonic component
which is also in agreement with observations.  The relative
normalization of halo, thick and thin disk surface mass densities in
the galactic plane \citep{san87} gives a proportion 1: 22: 200, which
implies that the halo surface mass density must be  1/100 of that of the disk
component. This result is in agreement with the ratio obtained by the
multiphase model \citep{fer92,pardi95}.

This collapse time scale also includes dynamical effects 
 and is, therefore, longer than a simple
free-fall time ( $\leq 1$ Gyr for the MWG). By using this method to
compute the characteristic collapse time scale for spiral galaxies
with masses different from our own, we are taking into account the
gravitational effect, although we are neglecting dynamical effects
that may be different from those of the MWG.

\begin{figure}
\plotone{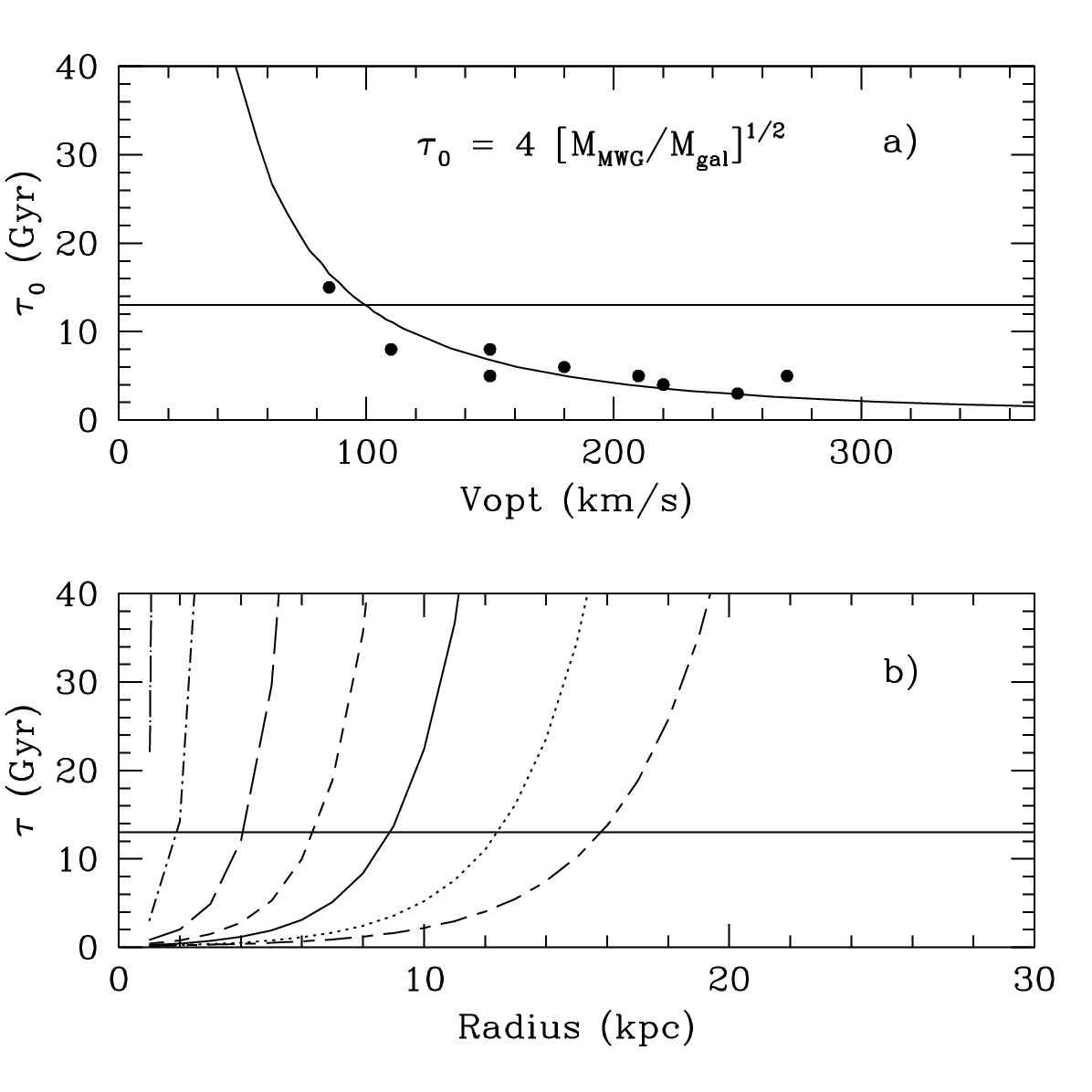}
\caption[]{a) Characteristic collapse timescale $\tau_{0}$ for each
galaxy according to the maximum rotation velocity.  b) Radial
distribution of the collapse times scales $\tau_{coll}(R)$.  Lines
have the same meaning than in Fig.~\ref{dis} The horizontal solid line
represents the assumed age of galaxies, 13 Gyr.}
\label{tcoll}
\end{figure}

The characteristic time scales for our models, $\tau_{0}$, are shown
in Fig.~\ref{tcoll}a) vs the rotation velocity at the optical radius
for each galaxy. Solid dots are the values used in our previous models
for individual spiral galaxies \citep{fer94,mol96,mol99}.  $\tau_{0}$
is defined for every galaxy as being that corresponding to the region
located at R$_{0}$ defined above and given in Column (7) of
Table~\ref{grid_m}.

An important consequence of the hypothesis linking the collapse time
scale with the radial distribution of the total mass, is that low mass
galaxies take more time to form their disks, in apparent contradiction
with the standard hierarchical picture of galaxy formation.  We are in
agreement with \citet{boi00} that this characteristic is, however,
essential to reproduce most observational features along the Hubble
Sequence, as the metallicity-magnitude relations or the colors of
disks.

It is evident that, following the same relation between collapse time
scale and mass, this collapse time scale must be variable with
galactocentric radius. If we assume that the total mass surface
density has an exponential form as the surface brightness, the
collapse time scale required to obtain that shape should depend on
radius through an exponential function, which increases with a scale
length $ \lambda_{D} \propto Re$ (where Re is the corresponding scale
length of the surface brightness radial distribution). We must bear in
mind, however, that the surface brightness distribution is the final
result of the combination of both the collapse and the star formation
processes, and therefore the collapse time scale may have in principle
a different dependence on radius than the surface brightness itself.
Nevertheless, for simplicity, we assume the collapse time scale as an
exponential function variable along the radius with a scale length
$\lambda_{D}$ which we assume equal to the $R_{D}/2$ given by PSS96:

\begin{equation}
\tau_{coll}(\rm R)=\tau_{0}\exp{((\rm R-R_{0})/\lambda_{D})}
\label{tau_r}
\end{equation}

This means that the scale length decreases for the later type of
galaxies and is larger for the earlier ones. This is in
agreement with observations from \citet[][see their Fig.~4]{vau}, and
has been more recently found by \citet{grah01} --but also see Fig. 2
from \citet{guz96}.

The radial dependence of the infall rate is not imposed {\sl a priori}
in our scenario: it is consequence of the gravitational law and of the
total mass distribution in the protogalaxy. The physical meaning is
clear: galaxies begin to form their inner regions before the outer
ones in a classical inside-out scheme. This halo-disk connection is
crucial for the understanding of the evolution of a galaxy from 
early times, the inside-out scenario resulting essential to reproduce
the radial gradient of abundances \citep{por99,boi00}.  In
fact, when the dynamical equations are also taken into account in a
chemo-dynamical model \citep{sam97}, this scenario is produced
naturally; thus our model yields results in good agreement with more
sophisticated models.

The radial variation of the collapse time scale for each galaxy,
calculated with equation (\ref{tau_r}) is shown in Fig.\ref{tcoll}b),
where we also draw a solid line at the 13 Gyr, the assumed age of
galaxies. If the collapse time scale is larger than this value, there
is not enough time for all the gas to fall onto the disk: only a small
part of it has moved from the halo to the equatorial plane and the
disk formation is not yet complete. This might explain the data from
\citet{san01}, who have found an extended component of H\,{\sc i},
different from the cold disk, located in the halo, rotating more
slowly than the disk and with radial inward motion.

In the various regions of the disk or bulge, and the halo, which treat
separately, we allow for different phases of matter aggregation:
diffuse gas ({\sl g}), clouds ({\sl c}, except in the halo), low-mass
($s_{1}, m < 4 M_{\odot}$) and massive stars ($s_{2}, m \ge 4
M_{\odot}$), and remnants. This border mass is related to
nucleosynthesis prescriptions: stars with masses lower than 4
M$_{\odot}$ only produce light elements, and do not contribute to the
interstellar medium enrichment.
\footnote{On the other hand, this discrimination in two groups, less
and more massive than 4 M$_{\odot}$, allow a very easy comparison of
our resulting metallicity distribution with the observed one, based in
G-dwarf low mass stars.}

The mass in the different phases of each region changes by the
following conversion processes:

\begin{enumerate}
\item Star formation by the gas-spontaneous fragmentation in the halo:
$\propto K g_{H}^{1.5}$ ( a Schmidt law with $n= 1.5$)
\item Cloud formation in the disk from diffuse gas: $\propto \mu
g^{1.5}_{D}$
\item Star formation in the disk from cloud-cloud collisions: $\propto
H c^{2}$
\item Induced Star formation in the disk {\sl via} massive star-cloud
interactions: $\propto a c s_{2}$
\item Diffuse gas restitution from these cloud and star formation
processes
\end{enumerate}

where $K$, $\mu$, $H$ and $a$ , besides the parameter $f$ or
$\tau_{0}$ already defined, are the parameters of the model.

Since the number of parameters is high, we have followed a precise
strategy to reduce to a minimum their degree of freedom.  Actually,
not all input parameters in our models can be considered as free. The
parameter $f$ is the inverse of the collapse time scale, which, in
turn, is defined by each mass radial distribution, as we have
explained above.

The other parameters, ($K$, $\mu$, $H$ and $a$) are calculated for
each radial region from the equations given in \citet{fer94}, which
give their dependence on the region volume
\footnote{The volume of the disk is calculated with a scale
height of 0.2 kpc for all galaxies} through proportionality factors or
{\sl efficiencies}. These efficiencies are the probabilities of
cloud formation, $\epsilon_{\mu}$, of cloud---cloud collision,
$\epsilon_{H}$, of the interaction of massive stars with clouds,
$\epsilon_{a}$ in the disk, and the efficiency to form stars in the
halo, $\epsilon_{K}$, and are characteristic of each spiral galaxy.

The term associated to the induced star formation describes a local
process and, as a result, its coefficient $\epsilon_{a}$ is considered
independent of both position and morphological type. The term
$\epsilon_{K}$ is also assumed constant for all halos, thus being
independent of morphological type \footnote{ The volume of the
halo for each concentric region, which has also an effect on the value
of the parameter K, is computed through the expression: $\rm V_{halo}(R)
= 2 R_{halo} R \sqrt{(1-\frac{R}{R_{halo}})^{2}}$.  Since $R_{halo}=2.5
R_{opt}$, $\rm V_{halo}$ takes a different value for each mass radial
distribution, but does not change with morphological type.}. Both
efficiencies take therefore the same values already used in our
previous model for the MWG, for all our 440 models.

The fact that galaxies with the same gravitational potential or
mass and different morphological type or appearance exist, implies
that the evolution of a galaxy does not depend solely on gravitation,
but also on certain dynamical conditions. These conditions cannot be
taken into account, obviously, in a simple chemical evolution model,
but may change the evolution of a galaxy, (mostly the star formation
rate through the temperature variations). They are included in our
efficiencies to form molecular clouds and stars, $\epsilon_{\mu}$ and
$\epsilon_{H}$, which are allowed to change from one galaxy to another.

Therefore, we assume that $\epsilon_{\mu}$ and $\epsilon_{H}$
vary between 0 and 1. Ten different values in this range are taken
for each radial mass distribution, in order to reproduce different
rates of evolution. We thus obtain all possible combinations of
collapse time scale with star and molecular cloud formation
efficiencies.

\begin{figure}
\plotone{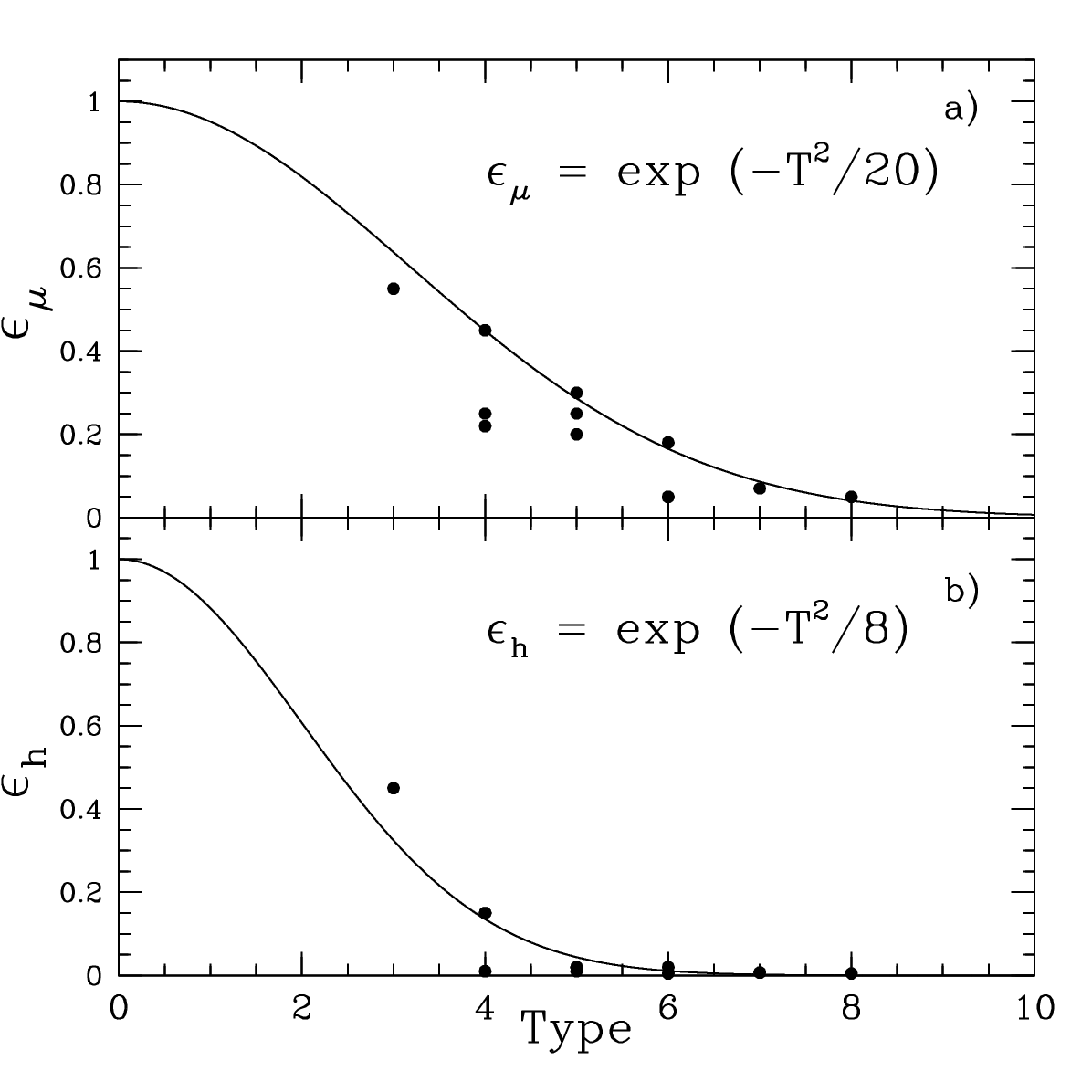}
\caption[]{Dependence of the efficiencies,$\epsilon_{\mu}$ and 
$\epsilon_{H}$ on morphological type T. The line is the fit
performed to the values used in our previous models, represented by
solid dots}
\label{efi}
\end{figure}

On the other hand, we know by our previous works that these
efficiencies change with galaxy morphological type. Thus, as we can
see in Fig.~\ref{efi}, the values used for some spiral galaxies in
\citet{mol96,mol99}, shown as full dots in both panels, depend on the
morphological type of these galaxies. In fact, this dependence was
already found in \citet{fer88,gal89}, where, these authors quantified
the efficiencies to form molecular clouds and the frequency of
cloud-cloud collisions, finding that a variation of 10 in the
parameters H and $\mu$ is needed when the Hubble type changes from one
stage to the next (Sa to Sb, etc..). They based their work in
the particle simulations of a clumpy interstellar medium from
\cite{rob84} and \citet{hau84}. 

Therefore, by taking into account the probability nature of our
efficiencies, we fit a probability function to the previous model
values:

\begin{equation}
\epsilon_{\mu}=exp{(-T^{2}/20)}
\end{equation}

\begin{equation}
\epsilon_{H}=exp{(-T^{2}/8)}
\end{equation}

These functions are shown as  solid lines in Fig.\ref{efi}, where the
points represent our previous model efficiencies.  With these
functions we  have computed 10 values for $\epsilon_{\mu}$ and
$\epsilon_{H}$, which we may relate to galaxy morphological type.  
We must make clear that this correspondence comes as a result from our
(previous) models.  However, we may have simply considered 10
different values computed with a probability function$ exp(-x^{2}/a)$
in the range [0,1].  We could have then checked the equivalence between
x and T {\sl a posteriori}.
 
Besides that, we have checked that these efficiencies take values
according to the estimations obtained from observed star forming
regions for some particular galaxies. Thus, the conversion of atomic
to molecular gas in NGC~ 224 is approximately 50\% \citep{lad88}, what
implies $\epsilon_{\mu}= 0.50$ in excellent agreement with the value
used for the model of this galaxy. In the same way, for NGC~ 598,
\citet{wil88} estimated the mean time to consume the molecular gas in
1.1 Gyr and for the total gas in 1.4 Gyr, that correspond to values
$\epsilon_{\mu} \sim 0.05 $ and $\epsilon_{H} \sim 0.01 $, similar to
those used in \citet{mol96}. Thus, our assumptions are supported by
observational studies.

{\sl Summarizing, only the characteristic collapse time scale,
depending on the total mass, and these two efficiencies,
$\epsilon_{\mu}$ and $\epsilon_{H}$ depending on morphological type,
are varied from galaxy to galaxy.} 

The enriched material proceeds from the restitution from dying stars,
considering their nucleosynthesis, their IMF (and the delayed
restitution) and their final fate, via a quiet evolution, or Type I
and II supernova explosions. Nucleosynthesis yields are taken from
\citet{woo95} for massive stars. For low mass and intermediate stars
we use the set of yields from \cite{ren81}.  For the type I supernova
explosion releases we take the model W7 from \cite{nom84}, as revised
by \citet{iwa99}.

\begin{figure}
\plotone{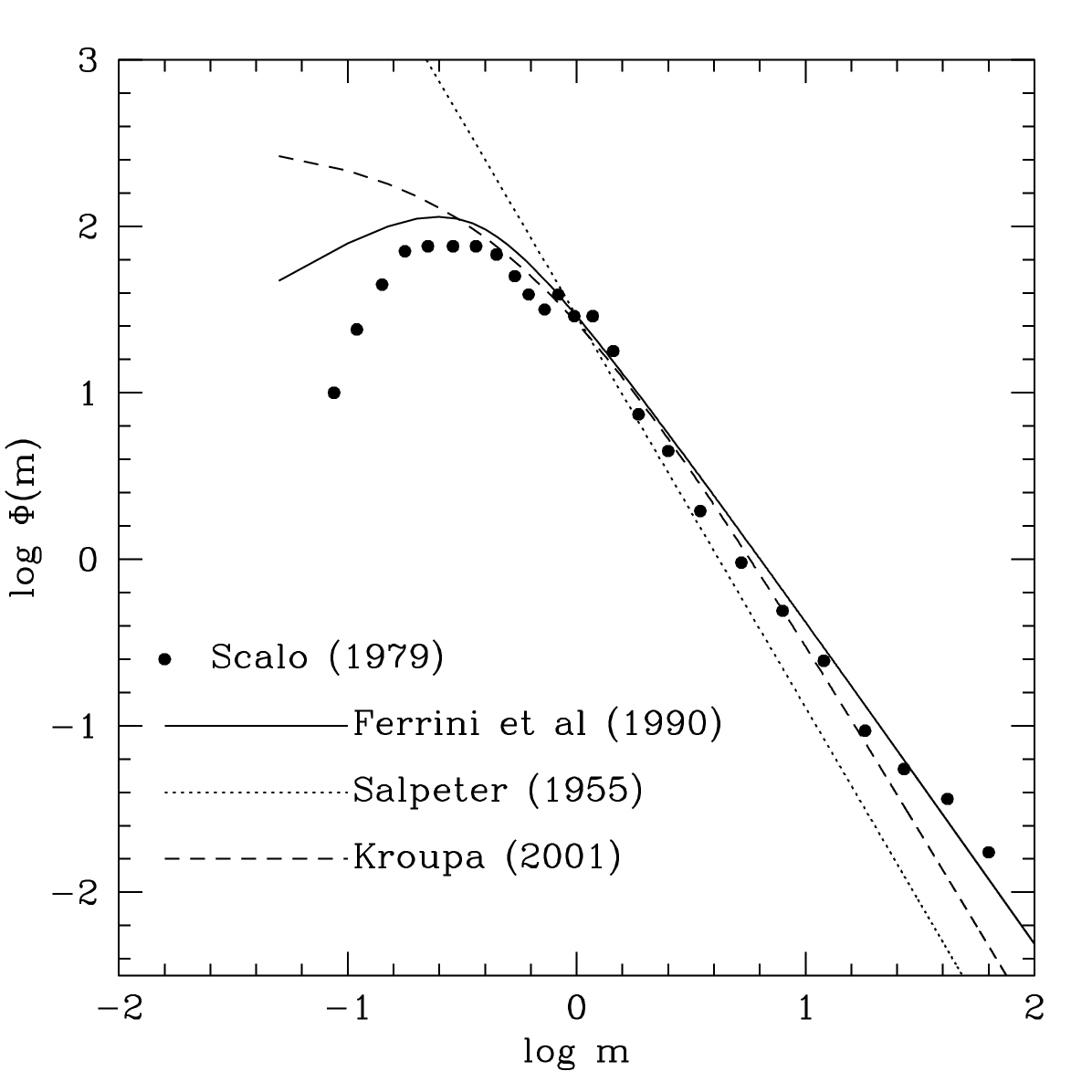}
\caption[]{The IMF from \citet{fer90}, solid line, compared to a
Salpeter law, dotted line, and that corresponding to \citet{kro01}, 
short-dashed line. Solid symbols correspond to \citet{sca86}.}
\label{imf}
\end{figure}

Most recent works support the idea that the IMF is practically
universal in space and constant in time \citep{wys97,sca98,mey00},
showing only local differences.  The adopted initial mass function
(IMF) is taken from \citet{fer90}, very similar to a Scalo's law and
in a good agreement with the most recent data from  \citet{kro01}, as
we can see in Fig.~\ref{imf}.

\section{Grid of Models: Presentation and Analysis}

With the model computed following the previous section, we obtain 440
different time evolutions, 10 for each radial distribution cited in
Table~\ref{dis}. The results corresponding to the mass of each region
and phase, the star formation rate and the supernova rates, for these
440 models are shown in Tables ~\ref{phases}, and ~\ref{sfrs}. The
elemental abundances for the disk are shown in Table~\ref{abundances}.
Here we only show, as an example, the results of the model
corresponding to the radial distribution number 19 and $T=5$ for the
first and last Gyr.  In Table ~\ref{phases} we list the time, in Gyr,
in Column (1), and the galactocentric distance R, in kpc, in Column
(2).  Columns (3) to (9) give the masses, in 10$^{9}
M_{\odot}$, in each region and phase: (3) the total mass in each
region, (4) the mass of the disk region, (5) the mass in the diffuse
gas phase, (6) the molecular gas, (7) the mass in low and intermediate
mass stars, (8) the mass in massive stars and (9) the mass in
remnants.

 In Table~\ref{sfrs}, we show for each time step in Gyr, column (1),
and radial distance, in kpc, column (2), the star formation rate, in
units of $M_{\odot} yr^{-1}$, in the disk and the halo regions in
Columns 3 and 4. The supernova rates, Types Ia and II, are in columns
(5) and (6), for the disk and (7) and (8) for the halo, respectively,
in units of 100 yr$^{-1}$.

The abundances in the disk for 15 elements are shown in
Table~\ref{abundances} also for each time in Gyr, column (1), and
galactocentric distance, column (2): H, D, $^{3}$He, $^{4}$He,
$^{12}$C, $^{13}$C, N, O, Ne, Si, S, Ca, and Fe are in columns from
(3) to (16), respectively. All of them are given by mass.

The complete tables with the complete time evolution from 0 to 13 Gyr,
with a time step of 0.5 Gyr, for the whole set of models may be
obtained from {http://cdsweb.u-strasbg.fr/Abstract.html}{http}, or
{http://pollux.ft.uam.es/astro/mercedes/grid}{http} or upon request to
authors.

\begin{deluxetable}{rrrrrrrrr}
\rotate
\tabletypesize{\footnotesize}
\tablecaption{Model Results corresponding to masses in each region and phase.
\label{phases}} 
\tablehead{
 \colhead{Time}  & \colhead{R}  &  
\colhead{Mtot} & \colhead{Mdisk} & \colhead{Mgas(HI)} 
& \colhead{Mgas(H$_{2}$)}& 
\colhead{Mstars(M$\rm < 4M_{\odot}$)}& 
\colhead{Mstars(M$\rm \ge 4M_{\odot}$)} &
  \colhead{Mremnants} \\
  \colhead{Gyr}  & \colhead{kpc} &
\colhead{(10$^{9}M_{\odot}$)} & \colhead{(10$^{9}M_{\odot}$)}
 &\colhead{(10$^{9}M_{\odot}$)}&\colhead{(10$^{9}M_{\odot}$)} 
 &\colhead{(10$^{9}M_{\odot}$)}&\colhead{(10$^{9}M_{\odot}$)} 
&\colhead{(10$^{9}M_{\odot}$)}  }
\startdata
   0.0 &10. & 0.44E+01 & 0.00E+00 & 0.10E-05 & 0.00E+00 & 0.00E+00 & 0.00E+00 & 0.00E+00 \\
   0.0 & 9. & 0.45E+01 & 0.00E+00 & 0.10E-05 & 0.00E+00 & 0.00E+00 & 0.00E+00 & 0.00E+00 \\
   0.0 & 8. & 0.45E+01 & 0.00E+00 & 0.10E-05 & 0.00E+00 & 0.00E+00 & 0.00E+00 & 0.00E+00 \\
   0.0 & 7. & 0.44E+01 & 0.00E+00 & 0.10E-05 & 0.00E+00 & 0.00E+00 & 0.00E+00 & 0.00E+00 \\
   0.0 & 6. & 0.43E+01 & 0.00E+00 & 0.10E-05 & 0.00E+00 & 0.00E+00 & 0.00E+00 & 0.00E+00 \\
   0.0 & 5. & 0.42E+01 & 0.00E+00 & 0.10E-05 & 0.00E+00 & 0.00E+00 & 0.00E+00 & 0.00E+00 \\
   0.0 & 4. & 0.39E+01 & 0.00E+00 & 0.10E-05 & 0.00E+00 & 0.00E+00 & 0.00E+00 & 0.00E+00 \\
   0.0 & 3. & 0.32E+01 & 0.00E+00 & 0.10E-05 & 0.00E+00 & 0.00E+00 & 0.00E+00 & 0.00E+00 \\
   0.0 & 2. & 0.21E+01 & 0.00E+00 & 0.10E-05 & 0.00E+00 & 0.00E+00 & 0.00E+00 & 0.00E+00 \\
   0.0 & 1. & 0.60E+00 & 0.00E+00 & 0.10E-05 & 0.00E+00 & 0.00E+00 & 0.00E+00 & 0.00E+00 \\
   0.5 &10. & 0.44E+01 & 0.60E-04 & 0.64E-04 & 0.80E-07 & 0.19E-14 & 0.72E-16 & 0.21E-16 \\
   0.5 & 9. & 0.45E+01 & 0.20E-03 & 0.20E-03 & 0.50E-06 & 0.78E-13 & 0.29E-14 & 0.85E-15 \\
   0.5 & 8. & 0.45E+01 & 0.61E-03 & 0.61E-03 & 0.32E-05 & 0.34E-11 & 0.13E-12 & 0.37E-13 \\
   0.5 & 7. & 0.44E+01 & 0.19E-02 & 0.18E-02 & 0.21E-04 & 0.15E-09 & 0.59E-11 & 0.17E-11 \\
   0.5 & 6. & 0.43E+01 & 0.57E-02 & 0.55E-02 & 0.14E-03 & 0.73E-08 & 0.27E-09 & 0.79E-10 \\
   0.5 & 5. & 0.42E+01 & 0.17E-01 & 0.16E-01 & 0.90E-03 & 0.35E-06 & 0.13E-07 & 0.38E-08 \\
   0.5 & 4. & 0.39E+01 & 0.48E-01 & 0.42E-01 & 0.56E-02 & 0.16E-04 & 0.57E-06 & 0.17E-06 \\
   0.5 & 3. & 0.32E+01 & 0.12E+00 & 0.92E-01 & 0.29E-01 & 0.54E-03 & 0.19E-04 & 0.62E-05 \\
   0.5 & 2. & 0.21E+01 & 0.23E+00 & 0.13E+00 & 0.95E-01 & 0.83E-02 & 0.26E-03 & 0.11E-03 \\
   0.5 & 1. & 0.60E+00 & 0.18E+00 & 0.64E-01 & 0.10E+00 & 0.18E-01 & 0.48E-03 & 0.26E-03 \\
   1.0 &10. & 0.44E+01 & 0.13E-03 & 0.13E-03 & 0.44E-06 & 0.12E-12 & 0.27E-14 & 0.19E-14 \\
   1.0 & 9. & 0.45E+01 & 0.39E-03 & 0.39E-03 & 0.28E-05 & 0.48E-11 & 0.11E-12 & 0.81E-13 \\
   1.0 & 8. & 0.45E+01 & 0.12E-02 & 0.12E-02 & 0.18E-04 & 0.21E-09 & 0.48E-11 & 0.35E-11 \\
   1.0 & 7. & 0.44E+01 & 0.37E-02 & 0.36E-02 & 0.12E-03 & 0.94E-08 & 0.22E-09 & 0.16E-09 \\
   1.0 & 6. & 0.43E+01 & 0.11E-01 & 0.11E-01 & 0.74E-03 & 0.43E-06 & 0.97E-08 & 0.73E-08 \\
   1.0 & 5. & 0.42E+01 & 0.34E-01 & 0.29E-01 & 0.46E-02 & 0.19E-04 & 0.41E-06 & 0.32E-06 \\
   1.0 & 4. & 0.39E+01 & 0.95E-01 & 0.69E-01 & 0.25E-01 & 0.67E-03 & 0.14E-04 & 0.12E-04 \\
   1.0 & 3. & 0.32E+01 & 0.24E+00 & 0.13E+00 & 0.93E-01 & 0.14E-01 & 0.24E-03 & 0.27E-03 \\
   1.0 & 2. & 0.21E+01 & 0.44E+00 & 0.16E+00 & 0.18E+00 & 0.92E-01 & 0.12E-02 & 0.22E-02 \\
   1.0 & 1. & 0.60E+00 & 0.31E+00 & 0.66E-01 & 0.14E+00 & 0.11E+00 & 0.99E-03 & 0.28E-02 \\
  \nodata &  \nodata &  \nodata &  \nodata &  \nodata &  \nodata &  \nodata &  \nodata &  \nodata \\
  12.0 &10. & 0.44E+01 & 0.14E-02 & 0.12E-02 & 0.18E-03 & 0.21E-06 & 0.51E-09 & 0.15E-07 \\
  12.0 & 9. & 0.45E+01 & 0.44E-02 & 0.34E-02 & 0.99E-03 & 0.73E-05 & 0.17E-07 & 0.53E-06 \\
  12.0 & 8. & 0.45E+01 & 0.13E-01 & 0.84E-02 & 0.48E-02 & 0.21E-03 & 0.43E-06 & 0.16E-04 \\
  12.0 & 7. & 0.44E+01 & 0.41E-01 & 0.20E-01 & 0.17E-01 & 0.37E-02 & 0.60E-05 & 0.32E-03 \\
  12.0 & 6. & 0.43E+01 & 0.12E+00 & 0.45E-01 & 0.41E-01 & 0.33E-01 & 0.37E-04 & 0.34E-02 \\
  12.0 & 5. & 0.42E+01 & 0.35E+00 & 0.86E-01 & 0.76E-01 & 0.17E+00 & 0.14E-03 & 0.20E-01 \\
  12.0 & 4. & 0.39E+01 & 0.92E+00 & 0.13E+00 & 0.12E+00 & 0.59E+00 & 0.36E-03 & 0.79E-01 \\
  12.0 & 3. & 0.32E+01 & 0.18E+01 & 0.13E+00 & 0.13E+00 & 0.13E+01 & 0.56E-03 & 0.20E+00 \\
  12.0 & 2. & 0.21E+01 & 0.19E+01 & 0.51E-01 & 0.85E-01 & 0.15E+01 & 0.27E-03 & 0.27E+00 \\
  12.0 & 1. & 0.60E+00 & 0.59E+00 & 0.57E-02 & 0.24E-01 & 0.46E+00 & 0.32E-04 & 0.97E-01 \\
  12.5 &10. & 0.44E+01 & 0.15E-02 & 0.13E-02 & 0.19E-03 & 0.26E-06 & 0.62E-09 & 0.19E-07 \\
  12.5 & 9. & 0.45E+01 & 0.45E-02 & 0.35E-02 & 0.11E-02 & 0.90E-05 & 0.20E-07 & 0.68E-06 \\
  12.5 & 8. & 0.45E+01 & 0.14E-01 & 0.86E-02 & 0.51E-02 & 0.25E-03 & 0.49E-06 & 0.20E-04 \\
  12.5 & 7. & 0.44E+01 & 0.43E-01 & 0.20E-01 & 0.18E-01 & 0.42E-02 & 0.64E-05 & 0.38E-03 \\
  12.5 & 6. & 0.43E+01 & 0.13E+00 & 0.46E-01 & 0.42E-01 & 0.36E-01 & 0.38E-04 & 0.38E-02 \\
  12.5 & 5. & 0.42E+01 & 0.37E+00 & 0.87E-01 & 0.76E-01 & 0.18E+00 & 0.14E-03 & 0.22E-01 \\
  12.5 & 4. & 0.39E+01 & 0.95E+00 & 0.13E+00 & 0.11E+00 & 0.62E+00 & 0.35E-03 & 0.85E-01 \\
  12.5 & 3. & 0.32E+01 & 0.18E+01 & 0.13E+00 & 0.13E+00 & 0.14E+01 & 0.54E-03 & 0.21E+00 \\
  12.5 & 2. & 0.21E+01 & 0.19E+01 & 0.49E-01 & 0.81E-01 & 0.15E+01 & 0.25E-03 & 0.28E+00 \\
  12.5 & 1. & 0.60E+00 & 0.59E+00 & 0.55E-02 & 0.23E-01 & 0.46E+00 & 0.29E-04 & 0.10E+00 \\
  13.0 &10. & 0.44E+01 & 0.15E-02 & 0.13E-02 & 0.21E-03 & 0.32E-06 & 0.74E-09 & 0.24E-07 \\
  13.0 & 9. & 0.45E+01 & 0.47E-02 & 0.35E-02 & 0.12E-02 & 0.11E-04 & 0.24E-07 & 0.85E-06 \\
  13.0 & 8. & 0.45E+01 & 0.15E-01 & 0.87E-02 & 0.55E-02 & 0.30E-03 & 0.56E-06 & 0.24E-04 \\
  13.0 & 7. & 0.44E+01 & 0.44E-01 & 0.21E-01 & 0.18E-01 & 0.48E-02 & 0.69E-05 & 0.44E-03 \\
  13.0 & 6. & 0.43E+01 & 0.13E+00 & 0.46E-01 & 0.43E-01 & 0.39E-01 & 0.39E-04 & 0.43E-02 \\
  13.0 & 5. & 0.42E+01 & 0.38E+00 & 0.87E-01 & 0.76E-01 & 0.19E+00 & 0.14E-03 & 0.24E-01 \\
  13.0 & 4. & 0.39E+01 & 0.98E+00 & 0.13E+00 & 0.11E+00 & 0.64E+00 & 0.35E-03 & 0.92E-01 \\
  13.0 & 3. & 0.32E+01 & 0.19E+01 & 0.12E+00 & 0.13E+00 & 0.14E+01 & 0.52E-03 & 0.23E+00 \\
  13.0 & 2. & 0.21E+01 & 0.19E+01 & 0.46E-01 & 0.78E-01 & 0.15E+01 & 0.23E-03 & 0.29E+00 \\
  13.0 & 1. & 0.60E+00 & 0.59E+00 & 0.53E-02 & 0.23E-01 & 0.46E+00 & 0.28E-04 & 0.10E+00 \\
\enddata
\end{deluxetable}

\begin{deluxetable}{rrrrrrrr}
\rotate
\tabletypesize{\footnotesize}
\tablecaption{Model Results: Star Formation Histories and Supernova Rates.
\label{sfrs}} 
\tablehead{ \colhead{Time}  & \colhead{R}  &  
\colhead{SFR(disk)} & \colhead{SFR(halo)} & \colhead{SN-Ia (disk)} 
& \colhead{SN-II (disk)}& 
\colhead{SN-Ia (halo)} & \colhead{SN-II(halo)}  \\
 \colhead{Gyr}  & \colhead{kpc}
&\colhead{M$_{\odot}yr^{-1}$} &\colhead{M$_{\odot}yr^{-1}$} 
& \colhead{$100 yr^{-1}$} & \colhead{$100 yr^{-1}$}& \colhead{$100 yr^{-1}$}
& \colhead{$100 yr^{-1}$} }
\startdata
   0.0 &10. & 0.00E+00 & 0.00E+00 & 0.00E+00 & 0.00E+00 & 0.00E+00 & 0.00E+00 \\
   0.0 & 9. & 0.00E+00 & 0.00E+00 & 0.00E+00 & 0.00E+00 & 0.00E+00 & 0.00E+00 \\
   0.0 & 8. & 0.00E+00 & 0.00E+00 & 0.00E+00 & 0.00E+00 & 0.00E+00 & 0.00E+00 \\
   0.0 & 7. & 0.00E+00 & 0.00E+00 & 0.00E+00 & 0.00E+00 & 0.00E+00 & 0.00E+00 \\
   0.0 & 6. & 0.00E+00 & 0.00E+00 & 0.00E+00 & 0.00E+00 & 0.00E+00 & 0.00E+00 \\
   0.0 & 5. & 0.00E+00 & 0.00E+00 & 0.00E+00 & 0.00E+00 & 0.00E+00 & 0.00E+00 \\
   0.0 & 4. & 0.00E+00 & 0.00E+00 & 0.00E+00 & 0.00E+00 & 0.00E+00 & 0.00E+00 \\
   0.0 & 3. & 0.00E+00 & 0.00E+00 & 0.00E+00 & 0.00E+00 & 0.00E+00 & 0.00E+00 \\
   0.0 & 2. & 0.00E+00 & 0.00E+00 & 0.00E+00 & 0.00E+00 & 0.00E+00 & 0.00E+00 \\
   0.0 & 1. & 0.00E+00 & 0.00E+00 & 0.00E+00 & 0.00E+00 & 0.00E+00 & 0.00E+00 \\
   0.5 &10. & 0.00E+00 & 0.74E-01 & 0.12E-14 & 0.68E-13 & 0.28E-01 & 0.23E+00 \\
   0.5 & 9. & 0.00E+00 & 0.75E-01 & 0.50E-13 & 0.28E-11 & 0.28E-01 & 0.23E+00 \\
   0.5 & 8. & 0.00E+00 & 0.77E-01 & 0.22E-11 & 0.12E-09 & 0.29E-01 & 0.24E+00 \\
   0.5 & 7. & 0.00E+00 & 0.79E-01 & 0.99E-10 & 0.55E-08 & 0.30E-01 & 0.24E+00 \\
   0.5 & 6. & 0.00E+00 & 0.82E-01 & 0.47E-08 & 0.26E-06 & 0.31E-01 & 0.25E+00 \\
   0.5 & 5. & 0.40E-05 & 0.83E-01 & 0.22E-06 & 0.12E-04 & 0.31E-01 & 0.25E+00 \\
   0.5 & 4. & 0.19E-03 & 0.80E-01 & 0.10E-04 & 0.53E-03 & 0.31E-01 & 0.24E+00 \\
   0.5 & 3. & 0.62E-02 & 0.67E-01 & 0.36E-03 & 0.17E-01 & 0.26E-01 & 0.20E+00 \\
   0.5 & 2. & 0.79E-01 & 0.37E-01 & 0.60E-02 & 0.23E+00 & 0.16E-01 & 0.11E+00 \\
   0.5 & 1. & 0.13E+00 & 0.56E-02 & 0.14E-01 & 0.38E+00 & 0.29E-02 & 0.17E-01 \\
   1.0 &10. & 0.00E+00 & 0.73E-01 & 0.89E-13 & 0.22E-11 & 0.51E-01 & 0.22E+00 \\
   1.0 & 9. & 0.00E+00 & 0.74E-01 & 0.37E-11 & 0.92E-10 & 0.52E-01 & 0.23E+00 \\
   1.0 & 8. & 0.00E+00 & 0.76E-01 & 0.16E-09 & 0.40E-08 & 0.53E-01 & 0.23E+00 \\
   1.0 & 7. & 0.00E+00 & 0.78E-01 & 0.73E-08 & 0.18E-06 & 0.55E-01 & 0.24E+00 \\
   1.0 & 6. & 0.30E-05 & 0.80E-01 & 0.33E-06 & 0.80E-05 & 0.56E-01 & 0.25E+00 \\
   1.0 & 5. & 0.12E-03 & 0.81E-01 & 0.15E-04 & 0.34E-03 & 0.57E-01 & 0.25E+00 \\
   1.0 & 4. & 0.39E-02 & 0.77E-01 & 0.53E-03 & 0.11E-01 & 0.55E-01 & 0.24E+00 \\
   1.0 & 3. & 0.62E-01 & 0.62E-01 & 0.11E-01 & 0.19E+00 & 0.46E-01 & 0.19E+00 \\
   1.0 & 2. & 0.28E+00 & 0.31E-01 & 0.77E-01 & 0.85E+00 & 0.26E-01 & 0.94E-01 \\
   1.0 & 1. & 0.23E+00 & 0.32E-02 & 0.88E-01 & 0.71E+00 & 0.40E-02 & 0.98E-02 \\
 \nodata &  \nodata &  \nodata &  \nodata &  \nodata &  \nodata &  \nodata &\\
  12.0 &10. & 0.00E+00 & 0.59E-01 & 0.13E-06 & 0.37E-06 & 0.14E+00 & 0.18E+00 \\
  12.0 & 9. & 0.40E-05 & 0.60E-01 & 0.45E-05 & 0.12E-04 & 0.14E+00 & 0.18E+00 \\
  12.0 & 8. & 0.10E-03 & 0.61E-01 & 0.12E-03 & 0.31E-03 & 0.15E+00 & 0.19E+00 \\
  12.0 & 7. & 0.14E-02 & 0.62E-01 & 0.20E-02 & 0.43E-02 & 0.15E+00 & 0.19E+00 \\
  12.0 & 6. & 0.87E-02 & 0.61E-01 & 0.16E-01 & 0.27E-01 & 0.15E+00 & 0.19E+00 \\
  12.0 & 5. & 0.32E-01 & 0.56E-01 & 0.67E-01 & 0.99E-01 & 0.14E+00 & 0.17E+00 \\
  12.0 & 4. & 0.83E-01 & 0.40E-01 & 0.19E+00 & 0.25E+00 & 0.11E+00 & 0.12E+00 \\
  12.0 & 3. & 0.13E+00 & 0.14E-01 & 0.35E+00 & 0.40E+00 & 0.48E-01 & 0.43E-01 \\
  12.0 & 2. & 0.63E-01 & 0.52E-03 & 0.24E+00 & 0.19E+00 & 0.56E-02 & 0.16E-02 \\
  12.0 & 1. & 0.73E-02 & 0.00E+00 & 0.42E-01 & 0.22E-01 & 0.19E-03 & 0.99E-07 \\
  12.5 &10. & 0.00E+00 & 0.59E-01 & 0.16E-06 & 0.44E-06 & 0.14E+00 & 0.18E+00 \\
  12.5 & 9. & 0.50E-05 & 0.59E-01 & 0.55E-05 & 0.14E-04 & 0.14E+00 & 0.18E+00 \\
  12.5 & 8. & 0.12E-03 & 0.60E-01 & 0.15E-03 & 0.35E-03 & 0.15E+00 & 0.18E+00 \\
  12.5 & 7. & 0.15E-02 & 0.61E-01 & 0.23E-02 & 0.46E-02 & 0.15E+00 & 0.19E+00 \\
  12.5 & 6. & 0.89E-02 & 0.60E-01 & 0.16E-01 & 0.27E-01 & 0.15E+00 & 0.18E+00 \\
  12.5 & 5. & 0.32E-01 & 0.55E-01 & 0.69E-01 & 0.99E-01 & 0.14E+00 & 0.17E+00 \\
  12.5 & 4. & 0.82E-01 & 0.39E-01 & 0.19E+00 & 0.25E+00 & 0.11E+00 & 0.12E+00 \\
  12.5 & 3. & 0.13E+00 & 0.13E-01 & 0.34E+00 & 0.38E+00 & 0.45E-01 & 0.40E-01 \\
  12.5 & 2. & 0.58E-01 & 0.44E-03 & 0.23E+00 & 0.18E+00 & 0.50E-02 & 0.13E-02 \\
  12.5 & 1. & 0.68E-02 & 0.00E+00 & 0.38E-01 & 0.21E-01 & 0.17E-03 & 0.70E-07 \\
  13.0 &10. & 0.00E+00 & 0.58E-01 & 0.20E-06 & 0.53E-06 & 0.14E+00 & 0.18E+00 \\
  13.0 & 9. & 0.60E-05 & 0.59E-01 & 0.66E-05 & 0.17E-04 & 0.14E+00 & 0.18E+00 \\
  13.0 & 8. & 0.13E-03 & 0.60E-01 & 0.17E-03 & 0.40E-03 & 0.15E+00 & 0.18E+00 \\
  13.0 & 7. & 0.16E-02 & 0.60E-01 & 0.25E-02 & 0.49E-02 & 0.15E+00 & 0.19E+00 \\
  13.0 & 6. & 0.92E-02 & 0.60E-01 & 0.17E-01 & 0.28E-01 & 0.15E+00 & 0.18E+00 \\
  13.0 & 5. & 0.32E-01 & 0.54E-01 & 0.70E-01 & 0.99E-01 & 0.14E+00 & 0.17E+00 \\
  13.0 & 4. & 0.81E-01 & 0.38E-01 & 0.19E+00 & 0.25E+00 & 0.10E+00 & 0.12E+00 \\
  13.0 & 3. & 0.12E+00 & 0.12E-01 & 0.33E+00 & 0.37E+00 & 0.43E-01 & 0.38E-01 \\
  13.0 & 2. & 0.54E-01 & 0.37E-03 & 0.21E+00 & 0.16E+00 & 0.44E-02 & 0.11E-02 \\
  13.0 & 1. & 0.64E-02 & 0.00E+00 & 0.35E-01 & 0.20E-01 & 0.16E-03 & 0.52E-07 \\
\enddata
\end{deluxetable}

\begin{deluxetable}{rrcccccccccccccc}
\rotate
\tabletypesize{\scriptsize}
\tablecaption{Model Results: Elemental Abundances.
\label{abundances}} 
\tablehead{ \colhead{Time}  & \colhead{R}  &  
\colhead{H} &  \colhead{D} & \colhead{$^{3}$He} &  \colhead{$^{4}$He}&
 \colhead{$^{12}$C}&  \colhead{$^{13}$C}& \colhead{$^{14}$N} &  
\colhead{O}& \colhead{Ne} & \colhead{Mg}& \colhead{Si}& \colhead{S}
&\colhead{Ca}& \colhead{Fe} }
\startdata
   0.0 &10. &0.770 & 0.70E-04 & 0.10E-04 &0.230 & 0.10E-09 & 0.10E-09 & 0.10E-09 & 0.10E-09 & 0.10E-09 & 0.10E-09 & 0.10E-09 & 0.10E-09 & 0.10E-09 & 0.10E-09 \\
   0.0 & 9. &0.770 & 0.70E-04 & 0.10E-04 &0.230 & 0.10E-09 & 0.10E-09 & 0.10E-09 & 0.10E-09 & 0.10E-09 & 0.10E-09 & 0.10E-09 & 0.10E-09 & 0.10E-09 & 0.10E-09 \\
   0.0 & 8. &0.770 & 0.70E-04 & 0.10E-04 &0.230 & 0.10E-09 & 0.10E-09 & 0.10E-09 & 0.10E-09 & 0.10E-09 & 0.10E-09 & 0.10E-09 & 0.10E-09 & 0.10E-09 & 0.10E-09 \\
   0.0 & 7. &0.770 & 0.70E-04 & 0.10E-04 &0.230 & 0.10E-09 & 0.10E-09 & 0.10E-09 & 0.10E-09 & 0.10E-09 & 0.10E-09 & 0.10E-09 & 0.10E-09 & 0.10E-09 & 0.10E-09 \\
   0.0 & 6. &0.770 & 0.70E-04 & 0.10E-04 &0.230 & 0.10E-09 & 0.10E-09 & 0.10E-09 & 0.10E-09 & 0.10E-09 & 0.10E-09 & 0.10E-09 & 0.10E-09 & 0.10E-09 & 0.10E-09 \\
   0.0 & 5. &0.770 & 0.70E-04 & 0.10E-04 &0.230 & 0.10E-09 & 0.10E-09 & 0.10E-09 & 0.10E-09 & 0.10E-09 & 0.10E-09 & 0.10E-09 & 0.10E-09 & 0.10E-09 & 0.10E-09 \\
   0.0 & 4. &0.770 & 0.70E-04 & 0.10E-04 &0.230 & 0.10E-09 & 0.10E-09 & 0.10E-09 & 0.10E-09 & 0.10E-09 & 0.10E-09 & 0.10E-09 & 0.10E-09 & 0.10E-09 & 0.10E-09 \\
   0.0 & 3. &0.770 & 0.70E-04 & 0.10E-04 &0.230 & 0.10E-09 & 0.10E-09 & 0.10E-09 & 0.10E-09 & 0.10E-09 & 0.10E-09 & 0.10E-09 & 0.10E-09 & 0.10E-09 & 0.10E-09 \\
   0.0 & 2. &0.770 & 0.70E-04 & 0.10E-04 &0.230 & 0.10E-09 & 0.10E-09 & 0.10E-09 & 0.10E-09 & 0.10E-09 & 0.10E-09 & 0.10E-09 & 0.10E-09 & 0.10E-09 & 0.10E-09 \\
   0.0 & 1. &0.770 & 0.70E-04 & 0.10E-04 &0.230 & 0.10E-09 & 0.10E-09 & 0.10E-09 & 0.10E-09 & 0.10E-09 & 0.10E-09 & 0.10E-09 & 0.10E-09 & 0.10E-09 & 0.10E-09 \\
   0.5 &10. &0.770 & 0.70E-04 & 0.10E-04 &0.230 & 0.59E-05 & 0.67E-07 & 0.14E-05 & 0.20E-04 & 0.40E-05 & 0.61E-06 & 0.65E-06 & 0.35E-06 & 0.52E-07 & 0.61E-06 \\
   0.5 & 9. &0.770 & 0.70E-04 & 0.10E-04 &0.230 & 0.60E-05 & 0.69E-07 & 0.14E-05 & 0.21E-04 & 0.41E-05 & 0.62E-06 & 0.66E-06 & 0.35E-06 & 0.53E-07 & 0.62E-06 \\
   0.5 & 8. &0.770 & 0.70E-04 & 0.10E-04 &0.230 & 0.62E-05 & 0.71E-07 & 0.15E-05 & 0.21E-04 & 0.42E-05 & 0.64E-06 & 0.68E-06 & 0.36E-06 & 0.55E-07 & 0.64E-06 \\
   0.5 & 7. &0.770 & 0.70E-04 & 0.10E-04 &0.230 & 0.64E-05 & 0.74E-07 & 0.15E-05 & 0.22E-04 & 0.43E-05 & 0.66E-06 & 0.71E-06 & 0.38E-06 & 0.57E-07 & 0.66E-06 \\
   0.5 & 6. &0.770 & 0.70E-04 & 0.10E-04 &0.230 & 0.67E-05 & 0.77E-07 & 0.16E-05 & 0.23E-04 & 0.46E-05 & 0.70E-06 & 0.74E-06 & 0.40E-06 & 0.60E-07 & 0.70E-06 \\
   0.5 & 5. &0.770 & 0.70E-04 & 0.10E-04 &0.230 & 0.71E-05 & 0.82E-07 & 0.17E-05 & 0.25E-04 & 0.48E-05 & 0.74E-06 & 0.79E-06 & 0.42E-06 & 0.63E-07 & 0.74E-06 \\
   0.5 & 4. &0.770 & 0.70E-04 & 0.10E-04 &0.230 & 0.77E-05 & 0.89E-07 & 0.19E-05 & 0.27E-04 & 0.53E-05 & 0.81E-06 & 0.86E-06 & 0.46E-06 & 0.69E-07 & 0.80E-06 \\
   0.5 & 3. &0.770 & 0.70E-04 & 0.10E-04 &0.230 & 0.11E-04 & 0.12E-06 & 0.27E-05 & 0.42E-04 & 0.83E-05 & 0.13E-05 & 0.13E-05 & 0.69E-06 & 0.10E-06 & 0.11E-05 \\
   0.5 & 2. &0.769 & 0.70E-04 & 0.10E-04 &0.230 & 0.36E-04 & 0.38E-06 & 0.92E-05 & 0.15E-03 & 0.30E-04 & 0.45E-05 & 0.46E-05 & 0.24E-05 & 0.35E-06 & 0.32E-05 \\
   0.5 & 1. &0.768 & 0.69E-04 & 0.10E-04 &0.231 & 0.95E-04 & 0.10E-05 & 0.24E-04 & 0.38E-03 & 0.75E-04 & 0.11E-04 & 0.12E-04 & 0.61E-05 & 0.89E-06 & 0.85E-05 \\
   1.0 &10. &0.770 & 0.70E-04 & 0.10E-04 &0.230 & 0.14E-04 & 0.17E-06 & 0.32E-05 & 0.41E-04 & 0.81E-05 & 0.13E-05 & 0.15E-05 & 0.80E-06 & 0.13E-06 & 0.20E-05 \\
   1.0 & 9. &0.770 & 0.70E-04 & 0.10E-04 &0.230 & 0.15E-04 & 0.17E-06 & 0.32E-05 & 0.42E-04 & 0.82E-05 & 0.13E-05 & 0.15E-05 & 0.82E-06 & 0.13E-06 & 0.20E-05 \\
   1.0 & 8. &0.770 & 0.70E-04 & 0.10E-04 &0.230 & 0.15E-04 & 0.18E-06 & 0.33E-05 & 0.43E-04 & 0.84E-05 & 0.13E-05 & 0.15E-05 & 0.84E-06 & 0.13E-06 & 0.21E-05 \\
   1.0 & 7. &0.770 & 0.70E-04 & 0.10E-04 &0.230 & 0.16E-04 & 0.18E-06 & 0.35E-05 & 0.45E-04 & 0.88E-05 & 0.14E-05 & 0.16E-05 & 0.87E-06 & 0.14E-06 & 0.22E-05 \\
   1.0 & 6. &0.770 & 0.70E-04 & 0.10E-04 &0.230 & 0.16E-04 & 0.19E-06 & 0.36E-05 & 0.47E-04 & 0.92E-05 & 0.14E-05 & 0.17E-05 & 0.92E-06 & 0.15E-06 & 0.23E-05 \\
   1.0 & 5. &0.770 & 0.70E-04 & 0.10E-04 &0.230 & 0.18E-04 & 0.21E-06 & 0.40E-05 & 0.52E-04 & 0.10E-04 & 0.16E-05 & 0.18E-05 & 0.10E-05 & 0.16E-06 & 0.24E-05 \\
   1.0 & 4. &0.770 & 0.70E-04 & 0.10E-04 &0.230 & 0.25E-04 & 0.29E-06 & 0.58E-05 & 0.78E-04 & 0.15E-04 & 0.24E-05 & 0.27E-05 & 0.15E-05 & 0.23E-06 & 0.33E-05 \\
   1.0 & 3. &0.769 & 0.70E-04 & 0.10E-04 &0.231 & 0.83E-04 & 0.96E-06 & 0.20E-04 & 0.27E-03 & 0.53E-04 & 0.82E-05 & 0.90E-05 & 0.48E-05 & 0.74E-06 & 0.97E-05 \\
   1.0 & 2. &0.765 & 0.68E-04 & 0.10E-04 &0.233 & 0.30E-03 & 0.36E-05 & 0.73E-04 & 0.91E-03 & 0.18E-03 & 0.28E-04 & 0.31E-04 & 0.17E-04 & 0.26E-05 & 0.36E-04 \\
   1.0 & 1. &0.762 & 0.67E-04 & 0.10E-04 &0.235 & 0.53E-03 & 0.67E-05 & 0.13E-03 & 0.15E-02 & 0.30E-03 & 0.46E-04 & 0.54E-04 & 0.30E-04 & 0.47E-05 & 0.72E-04 \\
  \nodata & \nodata &  \nodata &  \nodata &  \nodata &  \nodata &  \nodata &  \nodata &  \nodata &  \nodata &  \nodata &  \nodata &  \nodata &  \nodata &  \nodata &  \nodata \\
   12.0 &10. &0.767 & 0.68E-04 & 0.11E-04 &0.232 & 0.21E-03 & 0.23E-05 & 0.42E-04 & 0.48E-03 & 0.92E-04 & 0.16E-04 & 0.30E-04 & 0.18E-04 & 0.34E-05 & 0.89E-04 \\
  12.0 & 9. &0.767 & 0.68E-04 & 0.11E-04 &0.232 & 0.21E-03 & 0.24E-05 & 0.43E-04 & 0.49E-03 & 0.94E-04 & 0.16E-04 & 0.31E-04 & 0.18E-04 & 0.34E-05 & 0.91E-04 \\
  12.0 & 8. &0.767 & 0.68E-04 & 0.11E-04 &0.232 & 0.24E-03 & 0.27E-05 & 0.50E-04 & 0.56E-03 & 0.11E-03 & 0.18E-04 & 0.35E-04 & 0.21E-04 & 0.38E-05 & 0.10E-03 \\
  12.0 & 7. &0.765 & 0.67E-04 & 0.12E-04 &0.234 & 0.39E-03 & 0.46E-05 & 0.83E-04 & 0.93E-03 & 0.18E-03 & 0.30E-04 & 0.55E-04 & 0.33E-04 & 0.61E-05 & 0.16E-03 \\
  12.0 & 6. &0.759 & 0.63E-04 & 0.14E-04 &0.237 & 0.83E-03 & 0.10E-04 & 0.19E-03 & 0.19E-02 & 0.37E-03 & 0.64E-04 & 0.12E-03 & 0.71E-04 & 0.13E-04 & 0.35E-03 \\
  12.0 & 5. &0.749 & 0.58E-04 & 0.18E-04 &0.243 & 0.15E-02 & 0.20E-04 & 0.41E-03 & 0.34E-02 & 0.67E-03 & 0.12E-03 & 0.23E-03 & 0.14E-03 & 0.26E-04 & 0.70E-03 \\
  12.0 & 4. &0.741 & 0.52E-04 & 0.23E-04 &0.249 & 0.20E-02 & 0.32E-04 & 0.68E-03 & 0.47E-02 & 0.92E-03 & 0.16E-03 & 0.34E-03 & 0.21E-03 & 0.40E-04 & 0.11E-02 \\
  12.0 & 3. &0.735 & 0.48E-04 & 0.29E-04 &0.252 & 0.24E-02 & 0.41E-04 & 0.91E-03 & 0.54E-02 & 0.11E-02 & 0.19E-03 & 0.44E-03 & 0.27E-03 & 0.52E-04 & 0.15E-02 \\
  12.0 & 2. &0.726 & 0.38E-04 & 0.49E-04 &0.259 & 0.28E-02 & 0.58E-04 & 0.13E-02 & 0.63E-02 & 0.13E-02 & 0.24E-03 & 0.63E-03 & 0.39E-03 & 0.77E-04 & 0.23E-02 \\
  12.0 & 1. &0.703 & 0.16E-04 & 0.98E-04 &0.274 & 0.39E-02 & 0.93E-04 & 0.21E-02 & 0.88E-02 & 0.18E-02 & 0.36E-03 & 0.11E-02 & 0.66E-03 & 0.13E-03 & 0.41E-02 \\
  12.5 &10. &0.767 & 0.68E-04 & 0.11E-04 &0.232 & 0.21E-03 & 0.24E-05 & 0.44E-04 & 0.50E-03 & 0.95E-04 & 0.16E-04 & 0.31E-04 & 0.19E-04 & 0.35E-05 & 0.94E-04 \\
  12.5 & 9. &0.767 & 0.68E-04 & 0.11E-04 &0.232 & 0.22E-03 & 0.25E-05 & 0.45E-04 & 0.51E-03 & 0.98E-04 & 0.17E-04 & 0.32E-04 & 0.19E-04 & 0.36E-05 & 0.96E-04 \\
  12.5 & 8. &0.767 & 0.68E-04 & 0.11E-04 &0.232 & 0.25E-03 & 0.29E-05 & 0.52E-04 & 0.59E-03 & 0.11E-03 & 0.19E-04 & 0.37E-04 & 0.22E-04 & 0.41E-05 & 0.11E-03 \\
  12.5 & 7. &0.764 & 0.67E-04 & 0.12E-04 &0.234 & 0.42E-03 & 0.49E-05 & 0.90E-04 & 0.10E-02 & 0.19E-03 & 0.32E-04 & 0.60E-04 & 0.36E-04 & 0.66E-05 & 0.17E-03 \\
  12.5 & 6. &0.758 & 0.63E-04 & 0.14E-04 &0.238 & 0.88E-03 & 0.11E-04 & 0.21E-03 & 0.21E-02 & 0.40E-03 & 0.68E-04 & 0.13E-03 & 0.76E-04 & 0.14E-04 & 0.38E-03 \\
  12.5 & 5. &0.748 & 0.57E-04 & 0.18E-04 &0.244 & 0.15E-02 & 0.21E-04 & 0.43E-03 & 0.36E-02 & 0.70E-03 & 0.12E-03 & 0.24E-03 & 0.14E-03 & 0.27E-04 & 0.74E-03 \\
  12.5 & 4. &0.740 & 0.52E-04 & 0.24E-04 &0.249 & 0.21E-02 & 0.33E-04 & 0.71E-03 & 0.48E-02 & 0.94E-03 & 0.17E-03 & 0.36E-03 & 0.21E-03 & 0.41E-04 & 0.11E-02 \\
  12.5 & 3. &0.735 & 0.48E-04 & 0.31E-04 &0.253 & 0.24E-02 & 0.42E-04 & 0.93E-03 & 0.54E-02 & 0.11E-02 & 0.20E-03 & 0.45E-03 & 0.27E-03 & 0.53E-04 & 0.15E-02 \\
  12.5 & 2. &0.725 & 0.37E-04 & 0.53E-04 &0.259 & 0.29E-02 & 0.59E-04 & 0.13E-02 & 0.64E-02 & 0.13E-02 & 0.25E-03 & 0.65E-03 & 0.40E-03 & 0.80E-04 & 0.24E-02 \\
  12.5 & 1. &0.702 & 0.14E-04 & 0.10E-03 &0.275 & 0.40E-02 & 0.95E-04 & 0.21E-02 & 0.89E-02 & 0.18E-02 & 0.37E-03 & 0.11E-02 & 0.67E-03 & 0.13E-03 & 0.42E-02 \\
  13.0 &10. &0.767 & 0.68E-04 & 0.11E-04 &0.232 & 0.22E-03 & 0.25E-05 & 0.46E-04 & 0.52E-03 & 0.99E-04 & 0.17E-04 & 0.33E-04 & 0.20E-04 & 0.37E-05 & 0.98E-04 \\
  13.0 & 9. &0.767 & 0.68E-04 & 0.11E-04 &0.232 & 0.23E-03 & 0.26E-05 & 0.47E-04 & 0.53E-03 & 0.10E-03 & 0.18E-04 & 0.34E-04 & 0.20E-04 & 0.38E-05 & 0.10E-03 \\
  13.0 & 8. &0.766 & 0.68E-04 & 0.11E-04 &0.232 & 0.27E-03 & 0.31E-05 & 0.55E-04 & 0.63E-03 & 0.12E-03 & 0.20E-04 & 0.39E-04 & 0.23E-04 & 0.43E-05 & 0.11E-03 \\
  13.0 & 7. &0.764 & 0.66E-04 & 0.12E-04 &0.234 & 0.45E-03 & 0.53E-05 & 0.97E-04 & 0.11E-02 & 0.20E-03 & 0.35E-04 & 0.65E-04 & 0.38E-04 & 0.71E-05 & 0.19E-03 \\
  13.0 & 6. &0.757 & 0.62E-04 & 0.14E-04 &0.238 & 0.94E-03 & 0.12E-04 & 0.22E-03 & 0.22E-02 & 0.42E-03 & 0.72E-04 & 0.14E-03 & 0.82E-04 & 0.15E-04 & 0.41E-03 \\
  13.0 & 5. &0.748 & 0.56E-04 & 0.19E-04 &0.245 & 0.16E-02 & 0.22E-04 & 0.46E-03 & 0.37E-02 & 0.72E-03 & 0.13E-03 & 0.25E-03 & 0.15E-03 & 0.29E-04 & 0.78E-03 \\
  13.0 & 4. &0.740 & 0.51E-04 & 0.24E-04 &0.250 & 0.21E-02 & 0.34E-04 & 0.73E-03 & 0.49E-02 & 0.96E-03 & 0.17E-03 & 0.37E-03 & 0.22E-03 & 0.42E-04 & 0.12E-02 \\
  13.0 & 3. &0.734 & 0.47E-04 & 0.32E-04 &0.253 & 0.24E-02 & 0.43E-04 & 0.95E-03 & 0.55E-02 & 0.11E-02 & 0.20E-03 & 0.46E-03 & 0.28E-03 & 0.54E-04 & 0.16E-02 \\
  13.0 & 2. &0.724 & 0.36E-04 & 0.56E-04 &0.260 & 0.29E-02 & 0.61E-04 & 0.14E-02 & 0.65E-02 & 0.13E-02 & 0.25E-03 & 0.67E-03 & 0.41E-03 & 0.82E-04 & 0.25E-02 \\
  13.0 & 1. &0.700 & 0.13E-04 & 0.11E-03 &0.276 & 0.40E-02 & 0.97E-04 & 0.22E-02 & 0.89E-02 & 0.18E-02 & 0.37E-03 & 0.11E-02 & 0.68E-03 & 0.14E-03 & 0.42E-02 \\
\enddata
\end{deluxetable}

We now analyze the general results obtained with this bi-parametric
grid of models, in particular the effect of the total mass and/or the
morphological type on the rate of evolution. In order to do this, we
represent them in four figures, Fig.~\ref{hi}, ~\ref{h2},~\ref{oh} and
~\ref{sfr}, which we will analyze in the following subsections. In
each one of them, we show 3 panels, corresponding to 3 different
values of $\lambda$, i.e. different maximum rotation velocities and/or
radial distribution of mas<ses   M(R). We have selected the values
corresponding to $\lambda=$0.03, 0.19 and 1.0, that means galaxies
with rotation velocities of 48, 100 and 200~ $\rm km s^{-1}$,
respectively, as representing typical examples of spiral galaxies.
For each panel we show the results for the 10 selected rates of
evolution, or equivalently 10 morphological types from $\rm T=1$, the
most evolved one, corresponding to the highest efficiency values
$\epsilon_{\mu}$ and $\epsilon_{H}$, to $\rm T=10$, the least evolved.

\subsection{Radial distributions of diffuse gas}

\begin{figure*}
\plotone{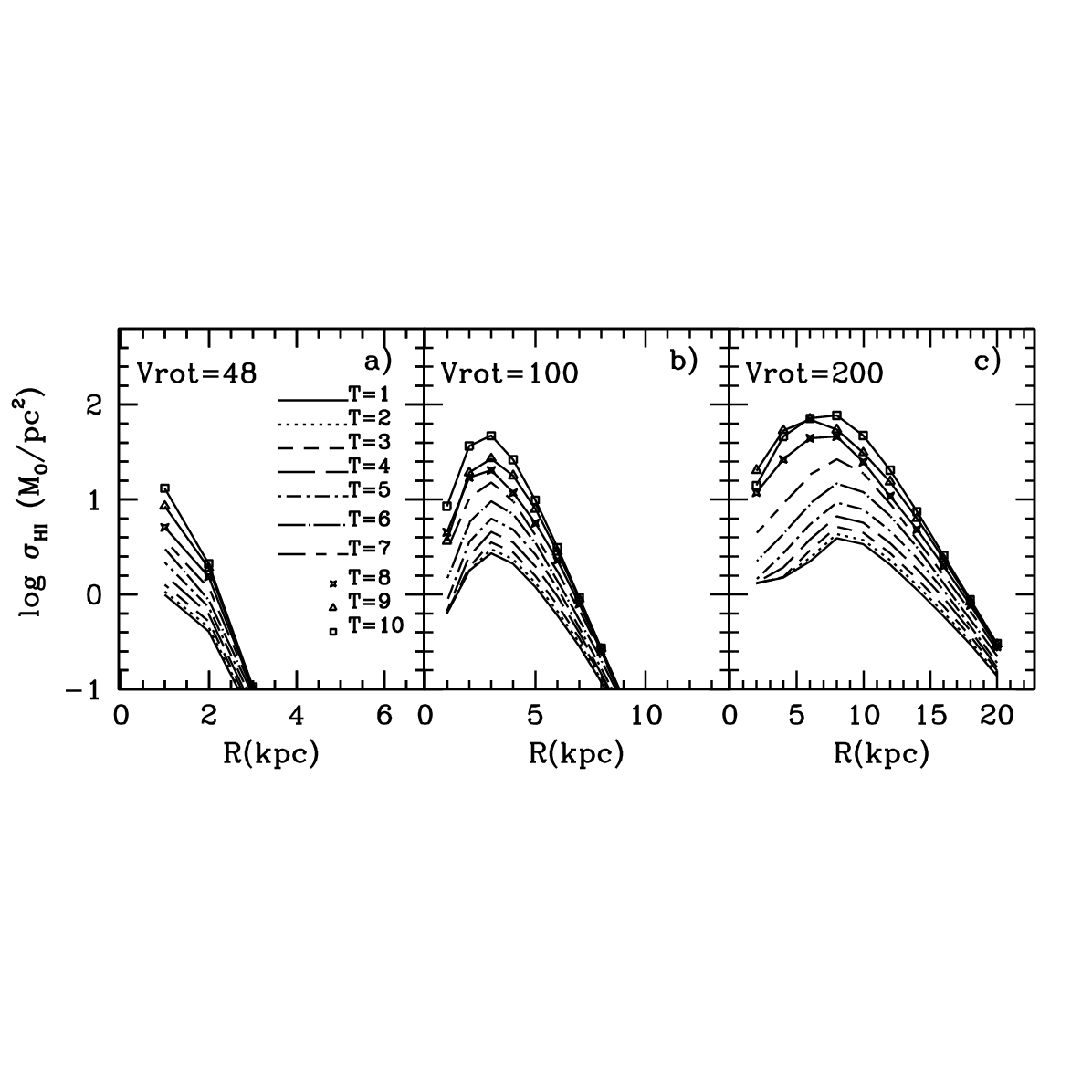}
\caption[]{Present epoch radial distributions of the logarithmic
surface density of the atomic gas for 3 different mass distributions
following the $Vrot$'s values from each panel. In each one of them 10
morphological types are represented following labels in panel a).}
\label{hi}
\end{figure*}

The radial distribution of atomic gas surface density is shown in
Fig.~\ref{hi}. We can see that the atomic gas surface density shows a
maximum in somewhere along the disk, as it is usually observed.  The
central value of this maximum depends on Hubble type: the earlier type
galaxies have smaller gas quantities and maximum values around 3-4
M$_{\odot}/pc^{2}$.  For intermediate types ($ 4 \leq T \leq 7$),
these maximum values rise to $\sim 5-8$ M$_{\odot}/pc^{2}$. For all
these morphological types the radial distributions are very similar
independently of their galactic mass, except for those corresponding
to $\lambda=0.03$ ($\rm Vrot= 48 km s^{-1}$) which show much lower
densities,except in the central region, for all morphological types.

The latest types (T $ > 7$) display a clear dependence on galactic
mass.  The maximum density values are always large, due to the small
efficiencies to form molecular clouds, which do not allow the
consumption of the diffuse gas. But this density is $\rm \sim 15
M_{\odot}/pc^{2}$ for $\lambda=0.03$ ($\rm Vrot= 48 km s^{-1}$) and
increases up to $\sim 30-40 M_{\odot}/pc^{2}$ for $\lambda=1.5$ ($\rm
Vrot= 248 km s^{-1}$).

In fact, a characteristic shown by all distributions is the similarity
among models of the same morphological type but different total mass,
when they are represented as a function of the normalized radius
$R/R_{opt}$. With the exception of the $\lambda=0.03$ model, all the
others show, for the same T, differences small enough as to simulate a
dispersion of the data.

The consequence of a shorter collapse time scale for the more massive
spirals is clearly seen: the maximum is located at radii further away
from the center due to the exhaustion of the diffuse gas in the inner
disk which moves the star formation outside. The smaller the galaxy
mass, the closer to the center is the maximum of the distribution,
which resembles an exponential, except for the inner region. In fact,
a shift in the maximum appears in each panel. In some cases, however,
the low values of the surface gas density are due to the fact that the
gas did not have enough time to fall completely onto the equatorial
disk. This effect is very clear for the model with $\lambda=0.03$
which shows a very steep distribution with densities lower than 15
$M_{\odot}/pc^{2}$.  In this case the gas shows a radial distribution
with a maximum at the center.

Therefore, besides the variations due to the differences in total
mass, which correspond to different collapse time-scales, a same total
mass may produce disks in different evolutionary states. Thus, a same
$\lambda=0.15 $ may result in a disk of 7-8 kpc and atomic gas
densities around 5 $M_{\odot}/pc^{2}$, or a disk of only 5 kpc with a
maximum density of 40 $M_{\odot}/pc^{2}$ in the region of 2 kpc.  On
the other hand, a galaxy with a large value of the total mass, may
show a high gas mass density, and a little disk or on the contrary, be
very evolved and therefore show no gas and a large stellar disk. The
first object could correspond to the low surface brightness galaxies,
while the last ones could be identified as the typical high surface
brightness spiral galaxies.

\subsection{Molecular gas radial distributions}

\begin{figure*}
\plotone{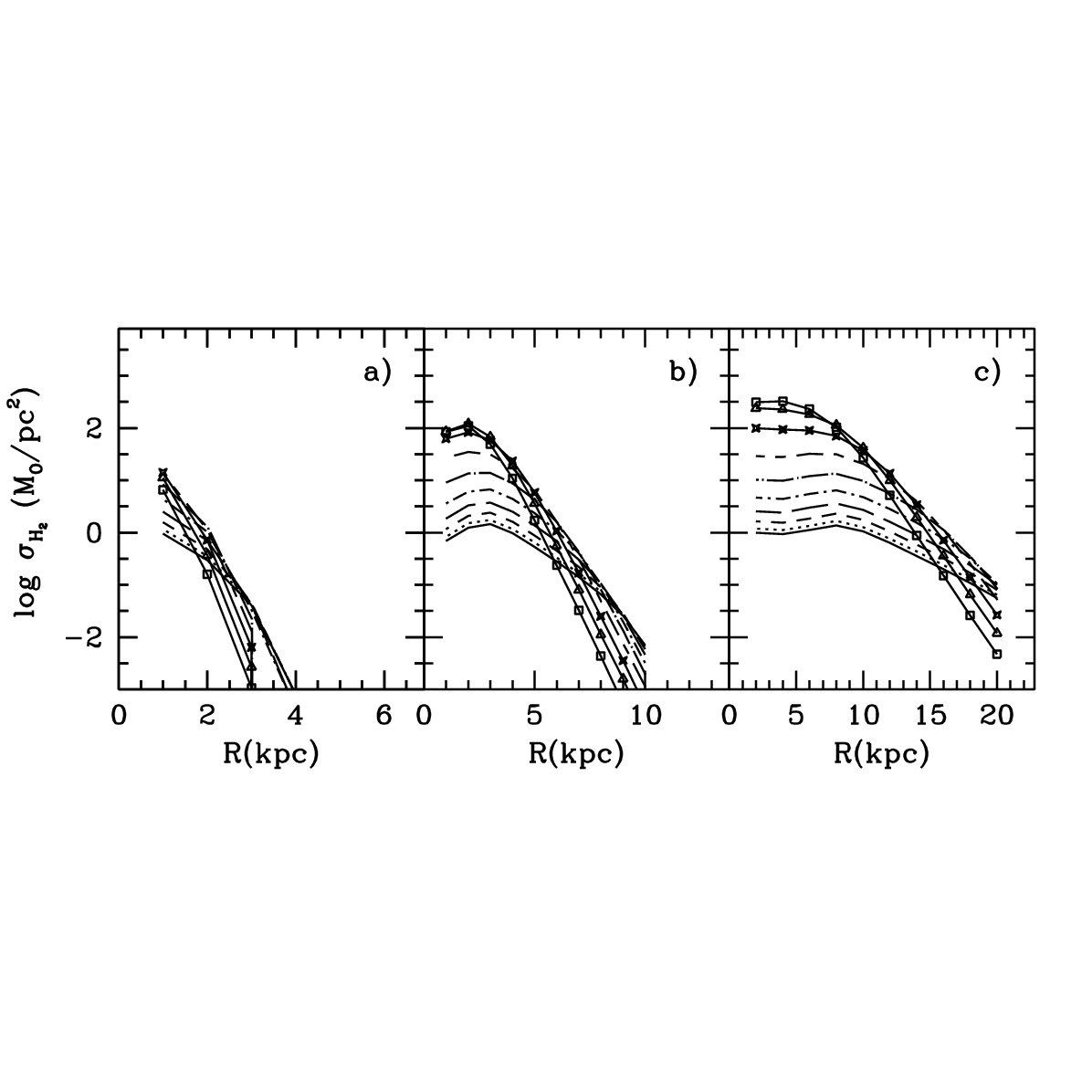}
\caption[]{Same as Fig.\ref{hi} for the molecular gas surface densities.
Symbols are the same than in that figure.}
\label{h2}
\end{figure*}

An important success of the multiphase models has been the ability to
reproduce the radial distributions for the atomic gas and the
molecular gas separately, which is possible due to the assumed star
formation prescription in two steps, allowing the formation of
molecular clouds prior to the appearance of stars. Since the
gas density in the disk depends only on the gravitational potential,
and the cloud formation rate depends both on the efficiency to form
clouds and the diffuse gas density, the molecular cloud density varies
with the gravitational potential, and also with the index T.

Another important consequence of this SFR law is that it takes into
account feedback mechanisms, even negative. If molecular clouds form
before stars, this implies a delay in the time of star formation.  The
massive stars formed induce, in turn, the creation of new stars. But,
at the same time these star formation processes also may destroy the
diffuse or molecular clouds, by preventing the total conversion of the
gas into stars and ejecting more gas once again into the ISM. In
particular, massive stars destroy the molecular clouds that surround
them, as \citet{par90} explained, due to the sensitivity of molecular
cloud condensation to the UV radiation. This mechanism restores gas to
the ISM, thus decreasing the star formation. Both regulating process
are included in our model. Neither heating
or cooling mechanisms for the cloud components are included in our
code.

The molecular gas shows an evolution similar to that of the diffuse
gas, but with a certain delay. This delay allows the existence of an
exponential function for a longer time, although in some evolved
galaxies $H_{2}$ is also consumed in the most central regions, thus
reproducing the so-called central {\sl hole} of the molecular gas
radial distribution. This is seen in Fig.\ref{h2} for
morphological types earlier than $T=5$, 6 or 7 (depending on the total
mass), for which the model lines turn over at the inner disk, which
corresponds to the regions located at the border between bulge and
disk.  Thus, the more evolved galaxies show a maximum in their radial
distribution of $\rm H_{2}$, which is always closer to the center than
that of the atomic gas distribution.

\begin{figure*}
\plotone{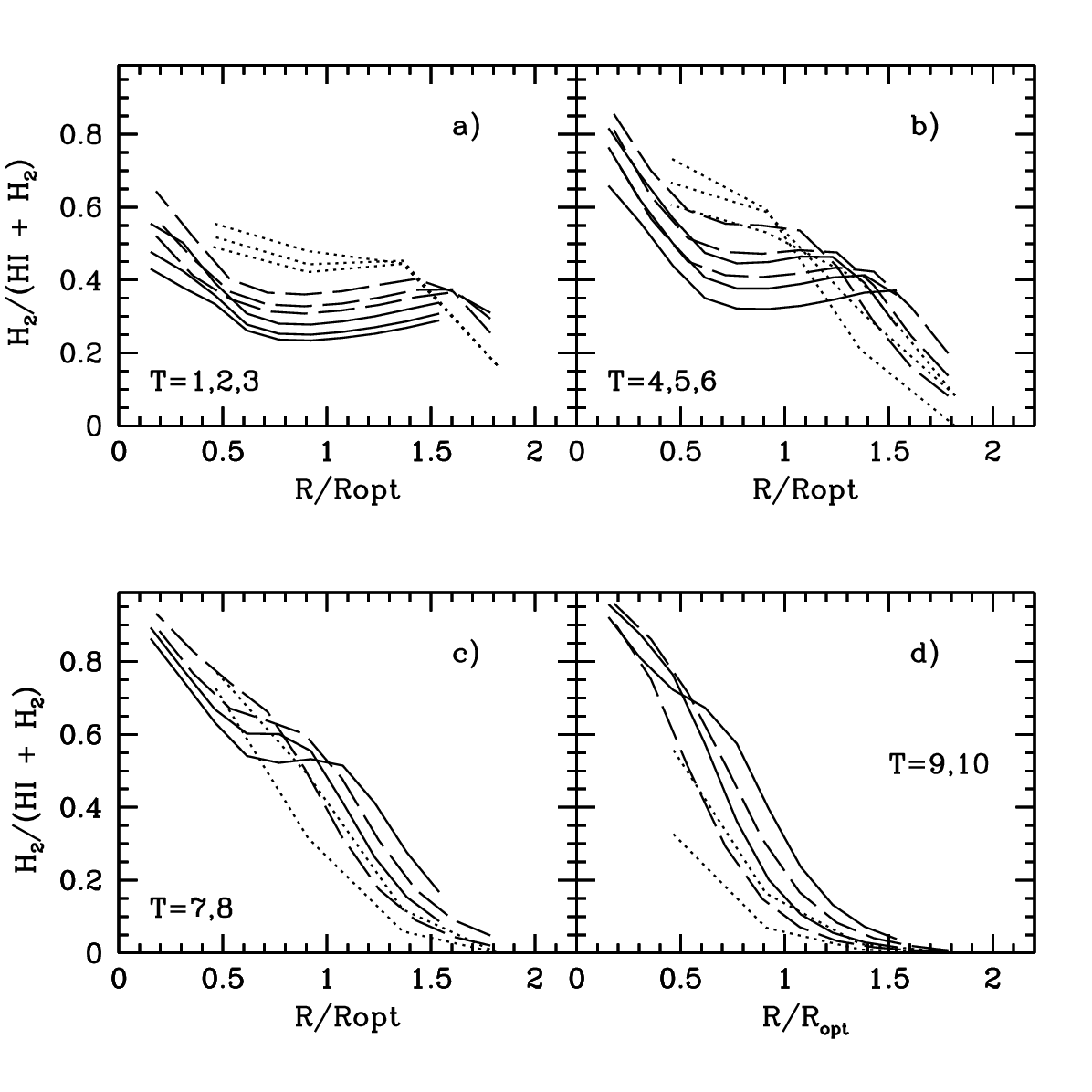}
\caption[]{Present epoch radial distributions of the  $H_{2}/M_{gas}$ ratio
for 3 different mass distributions in each panel. Labels have the same
meaning than in Fig.~\ref{dis}}
\label{h2_gtot}
\end{figure*}

The total quantity of gas in molecular form depends on two process:
the creation of molecular clouds, (that depends on the diffuse gas
mass available) and the process of conversion of the clouds into
stars.  Both effects have different rates for different type
T. Therefore large quantities of molecular gas are only found in
intermediate type galaxies: the earliest ones consume the molecular
gas very quickly, by decreasing the surface densities to values
smaller than 6 $M_{\odot}/pc^{2}$, as observed in NGC 224, while the
latest ones present rather long time scale to create clouds, due to
the low efficiencies and to the low amount of diffuse gas available.
This effect implies that the radial distributions do not have a
continuous behavior with T: for T$= 10$, the radial distributions are,
for some total masses, below those corresponding to earlier
morphological types, as T$=7$, 8, or 9.  This effect is shown in
Fig.~\ref{h2_gtot}, where we show, for each morphological type, the
distributions of the ratio of molecular to total mass gas
$H_{2}/M_{gas}$ {\sl vs} normalized radius $R/R{opt}$. In each panel
we show results for the different total mass values used in
previous figures. We put together the same morphological type models
giving similar  distributions. Thus we see that there are 4 kind of behavior:
The most evolved (T$=1 to 3$) models have ratios as low as $\sim 0.3$ for
all radial regions. For types 4 to 6 there exist models with a central
maximum and some others with this maximum located out the central region.
For types 7 and 8, the radial distributions show high values for the ratio
$H_{2}/M_{gas}$ for whole the disk, while for T$=10$ the ratio is small
except for the center where this is very high, almost 1.

Late-type galaxies show larger surface densities of
molecular than atomic gas because the efficiency to form stars from
molecular clouds is smaller that the efficiency to form these
clouds. That it, the conversion of diffuse to molecular gas occurs
more rapidly that the subsequent formation of stars. Thus these models
predict larger quantities of molecular gas for the less evolved
galaxies.

\subsection{The radial stellar disks profiles}

\begin{figure*}
\plotone{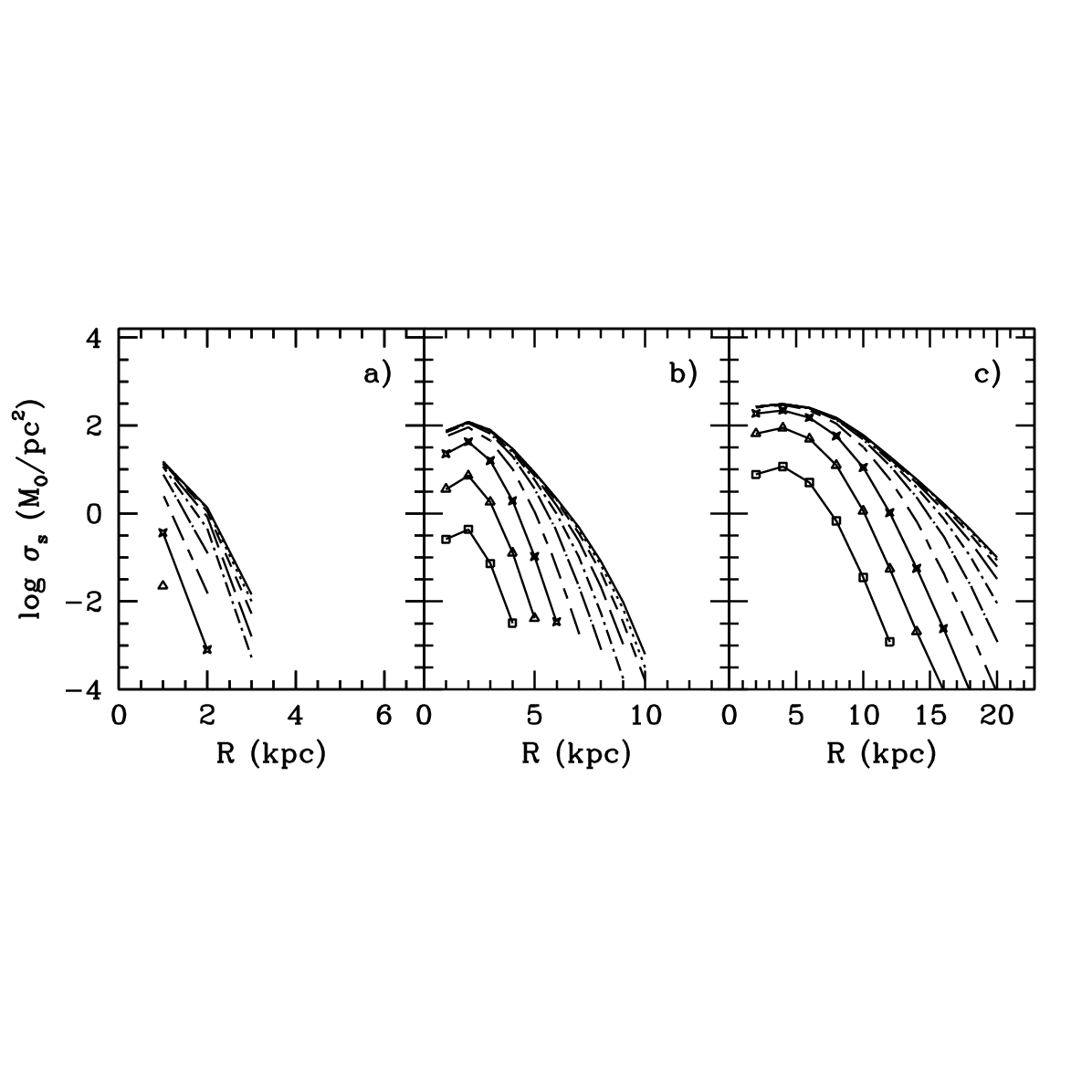}
\caption[]{Present epoch radial distributions of total mass surface 
density for 6 different mass distributions.
Ten different morphological type are shown in each  panel. Symbols are the
same than in Fig.\ref{hi}.}
\label{profiles}
\end{figure*}

The total mass converted into stars form out the stellar disk in each
galaxy.  These stellar disks are reproduced in Fig.\ref{profiles}
where we show in each panel the corresponding stellar surface density
radial distributions for a given radial distribution of total mass.

In this case the morphological type has less influence in the
resulting shape: the total mass of stars created is similar for all
types T $< 6$, although they are formed at different rates, that is
the resulting stellar populations have different mean ages. In the
earliest types, stars were created very rapidly, while in the latest
types, they formed later.  Therefore the radial distributions of
surface brightness results very similar for galaxies of all
morphological types for a given galactic total mass, but colors are
expected to be different, redder for the earlier type galaxies.

A very interesting result is that the central value is
practically the same, around 100 M$_{\odot}/pc^{2}$, for all rotation
curves and morphological types in agreement with Freeman's law. Only the less
evolved galaxies or the less massive disks show central densities
smaller than this value. We cannot compute a surface brightness only
with these models, but assuming a ratio $M/L=1 $ for the stellar populations,
this implies a surface luminosity density of 100 L$_{\odot}/pc^{2}$
and scale lengths in agreement with observed generic trends.
In any case, all information related to photometric quantities, and
this also applies to the disk scale lengths, must be computed through
the application of evolutionary synthesis models. These must be
calculated from the star formation histories resulting from the chemical
evolution models shown here, but this is out of the scope of this work
and will be adressed in a forthcoming work.

Only in models with T$> 7$ the stellar disks show a different
appearance: they look less massive, as corresponding to disks in the
process of formation.  This implies that the surface brightness is
lower than for the other types for a similar characteristic
total mass in the protogalaxy.

\begin{figure} 
\plotone{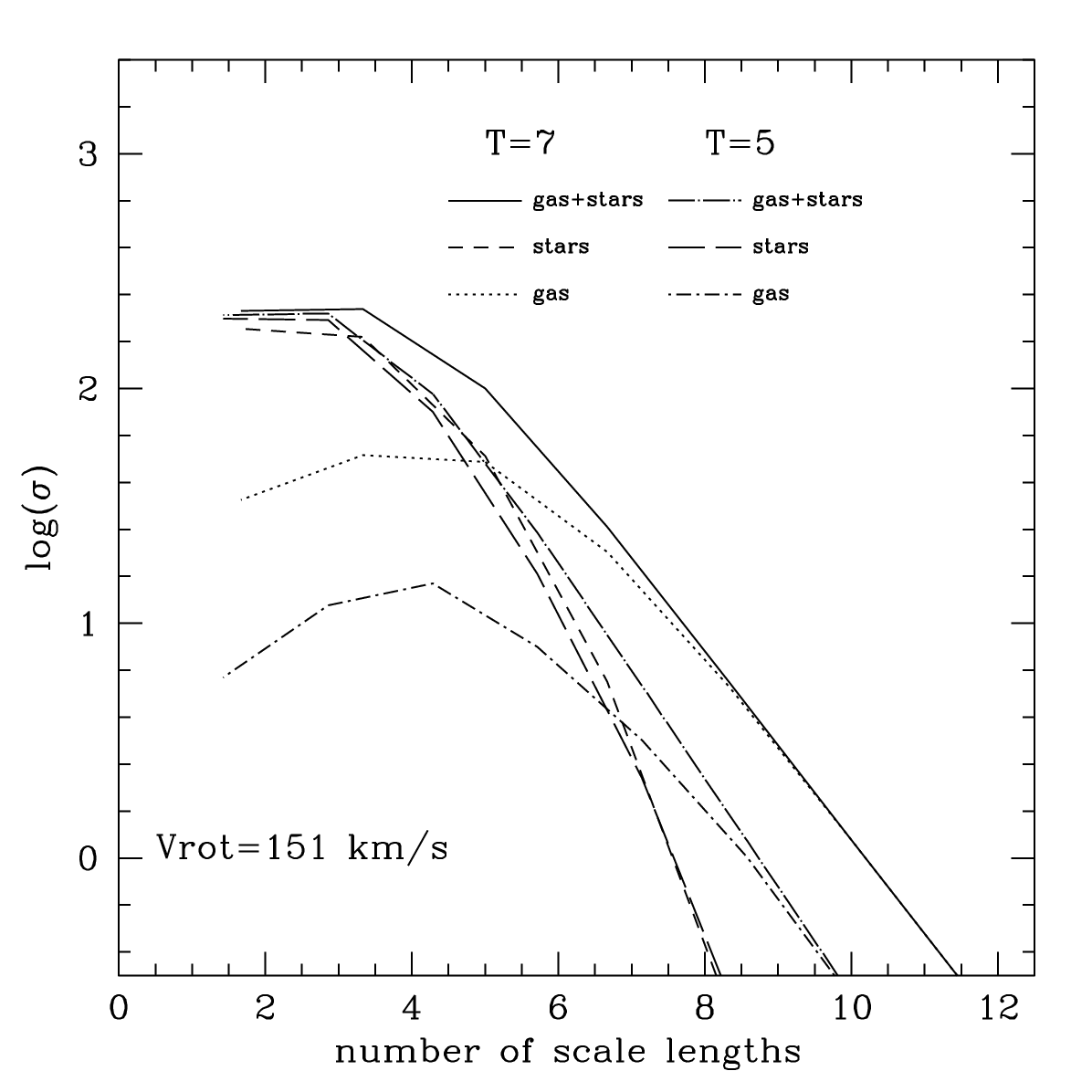}
\caption[]{Present epoch radial distributions of stars, gas and gas
plus stars, for 2 models with the same radial distribution of total
mass, corresponding to V$_{rot}=151 km s^{-1}$, and 2 different
morphological types, $\rm T=4$ and $\rm T=7$.}
\label{dos}
\end{figure}

In this way, we can reproduce the characteristics observed by
\citet{blok96} who show how two galaxies with very similar rotation
curves may, however, have very different radial distributions of gas
and stars. In Fig.\ref{dos} we represent the radial distribution of
stars, gas and gas plus stars predicted by 2 models computed with the
same radial distribution of total mass and different efficiencies
corresponding to morphological types 5 and 7. The X-axis is taken as
the measure of scale lengths obtained from the corresponding stellar
radial distribution shown in Fig.~\ref{profiles}.  The first model,
$T=7$, has a large amount of gas, still not consumed, and its stellar
profile differs strongly from that corresponding to its total mass.
We can see, however, that the model for $T=5$ shows a lower gas
density for the whole disk than that for $T=7$, due to its larger star
formation efficiency. Its stellar component has a cutoff radius
shorter than the gas but, except for this feature, it is very close to
the total mass distribution. We must realize that the $T=7$ model
looks more extended than type 5. This effect is due to the
representation on scale length units. Due to the steeper stellar
profile for the later type, which only produces stars at the center of
the disk, the scale length is smaller than that of the earlier type
model, which has a flat stellar radial distribution. Therefore, more
scale lengths are needed to represent the whole galaxy. Nevertheless,
the comparison of our Fig.\ref{dos} with Fig.2.  from \citet{blok96}
shows that it is extremely similar, thus proving that galaxies found
to have the same mass and different surface brightness may also be
explained with our models as resulting from different cloud
and star formation rates or different efficiencies. These efficiencies
include the dynamical information not treated explicitely in our
models, and therefore, the existence of two models with the same total
mass and different surface brightness may only be explained by
differences in the temperature and other thermodynamical conditions,
which would not depend on gravitational potential.

\subsection{The elemental abundances}

\begin{figure*}
\plotone{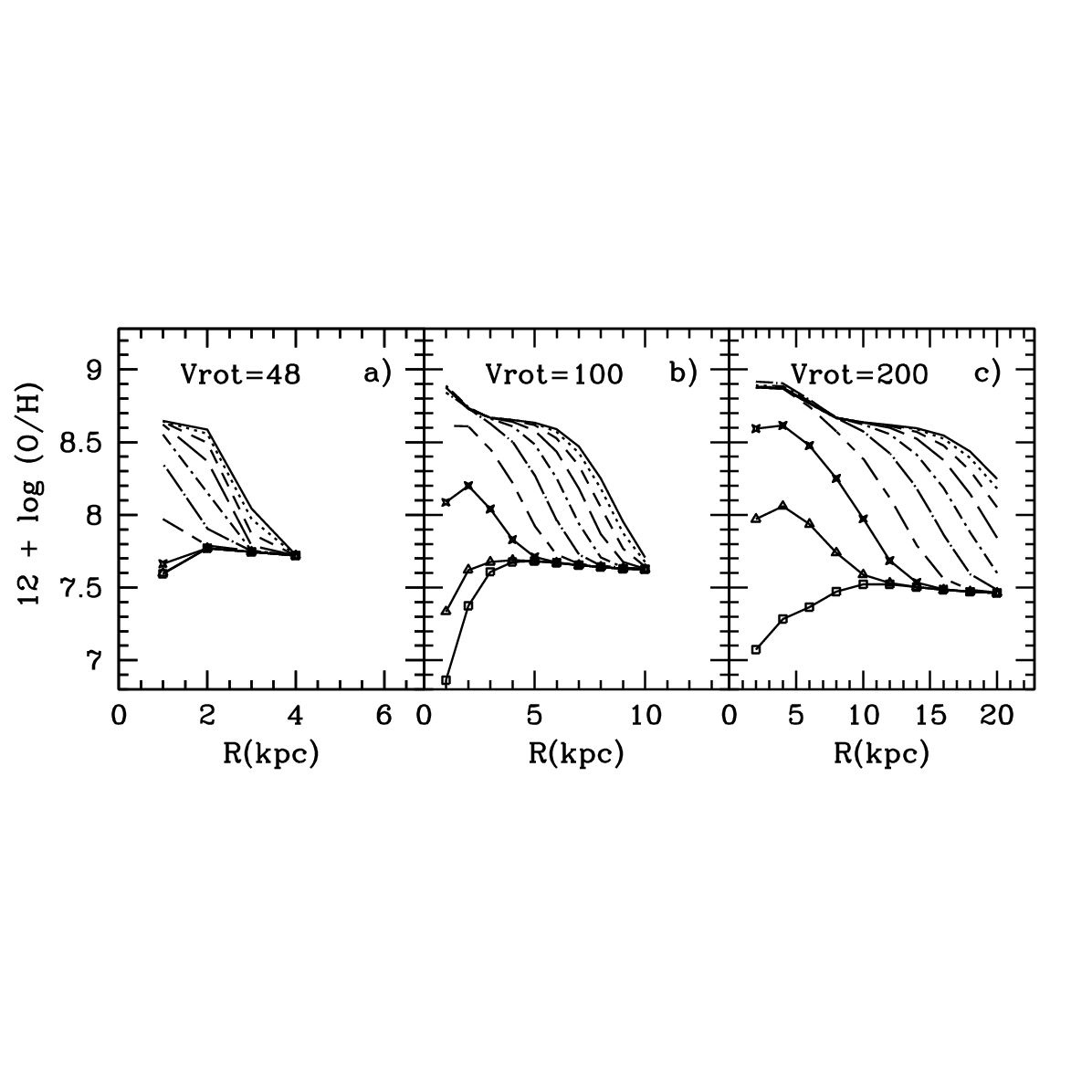}
\caption[]{Present epoch radial distributions  of
Oxygen abundance, $12+log (O/H)$, for 3 different mass distributions.
Ten different morphological type are shown in each  panel. Symbols are the
same than in Fig.\ref{hi}}
\label{oh}
\end{figure*}

One of the most important results of this grid of models refers to the
oxygen abundances, shown in Fig.~\ref{oh}. A radial gradient appears
for most of models. This is due to the different evolutionary rates
along radius: the inner regions evolve more rapidly that the outer
ones, thus steepening the radial gradient very soon for most
models. Then, the radial gradient flattens for the more massive and/or
earlier galaxies due to the rapid evolution which produces a large
quantity of elements, with the oxygen abundance reaching a saturation
level. This level is found to be around $12+log(O/H) \sim 9.0- 9.1$
dex. Observations in the inner disk of our Galaxy support this
statement \citep{sma01}.

Moreover, the larger the mass of the galaxy, the faster the effect: a
galaxy with V$_{rot}=100 km s^{-1}$ has a flat radial gradient for types
T$=$ 1 or 2, while a galaxy with V$_{rot}=200 km s^{-1}$, shows a flat
distribution for all types earlier than 4.  The less massive galaxies
maintain a steeper radial distribution of oxygen for almost all types,
with very similar values for the gradients.  Our models reproduce very
well the observed trend: the later the type of a galaxy, the steeper
its radial distribution of abundances.

Nevertheless, for any galaxy mass, {\sl the latest types $T> 7-8$ show
flat radial abundances distributions}. Thus, the late galaxies types show no
gradient, the intermediate types show steep gradients, and the early
galaxies have, once again, flat abundance radial distributions. The largest
values of radial gradients correspond to the intermediate types of
galaxies, with the limiting types varying according to  the total
mass of the galaxy. The more massive galaxies only show a significant
radial gradient for T$= 8$ and 9, while the less massive ones have a
flat gradient only for $\rm T =$9 or 10, the other types having very
pronounced radial gradients even for $T=1$.

In the early type galaxies, the characteristic efficiency
$\epsilon_{\mu}$ is high for all the disk, thus producing a high and
early star formation in all radial regions. In this case, the oxygen
abundance reaches very soon the saturation level or {\sl effective
yield}, flattening the radial gradient developed in the first times
of the evolution. The characteristic oxygen abundance, measured at
R$_{0}$, is higher for the more massive galaxies and lower for the
less massive ones. However, this correlation is not apparent when the
central abundance is used, due to the existence of this saturation
level in the oxygen abundance, which produces a flattening of the
radial gradient in the inner disk, even for intermediate type galaxies.

In fact, the oxygen abundance radial distribution shows
sometimes a bad fit to reported distributions in the central parts of
the disks, which give frequently abundances larger than 9.10 dex.
This absolute level of abundance is not reached in any case by the
models.  All the computations performed within the multiphase approach
reach a maximum $12 + log (O/H) \sim 9.10$ dex which no model can
exceed. It should be recalled, however, that all oxygen data yielding
values larger than 9.1 dex have been obtained in H{\sc ii} regions
where the electronic temperature could not be measured, and abundances
have been estimated through empirical calibrations which are very
uncertain in the high abundance regime. In fact, the shape of the
radial distribution of oxygen changes depending on the calibration
used to estimate them \citep[{\sl e.g.}][]{ken96}.  The suspicion that
these abundances are overestimated at least by a factor 0.2 dex is
very reasonable \citep{pil00,pil01,dia00,cas02}.

Therefore, the gradient behavior found in the models seems to be in
agreement with observations solving the apparent inconsistency in the
trends giving larger gradients for late types of galaxies and flatter
ones for the earliest ones while some irregulars show no gradient at
all, with very uniform abundances.

\subsection{Relative  Abundances} 

\begin{figure*}
\plotone{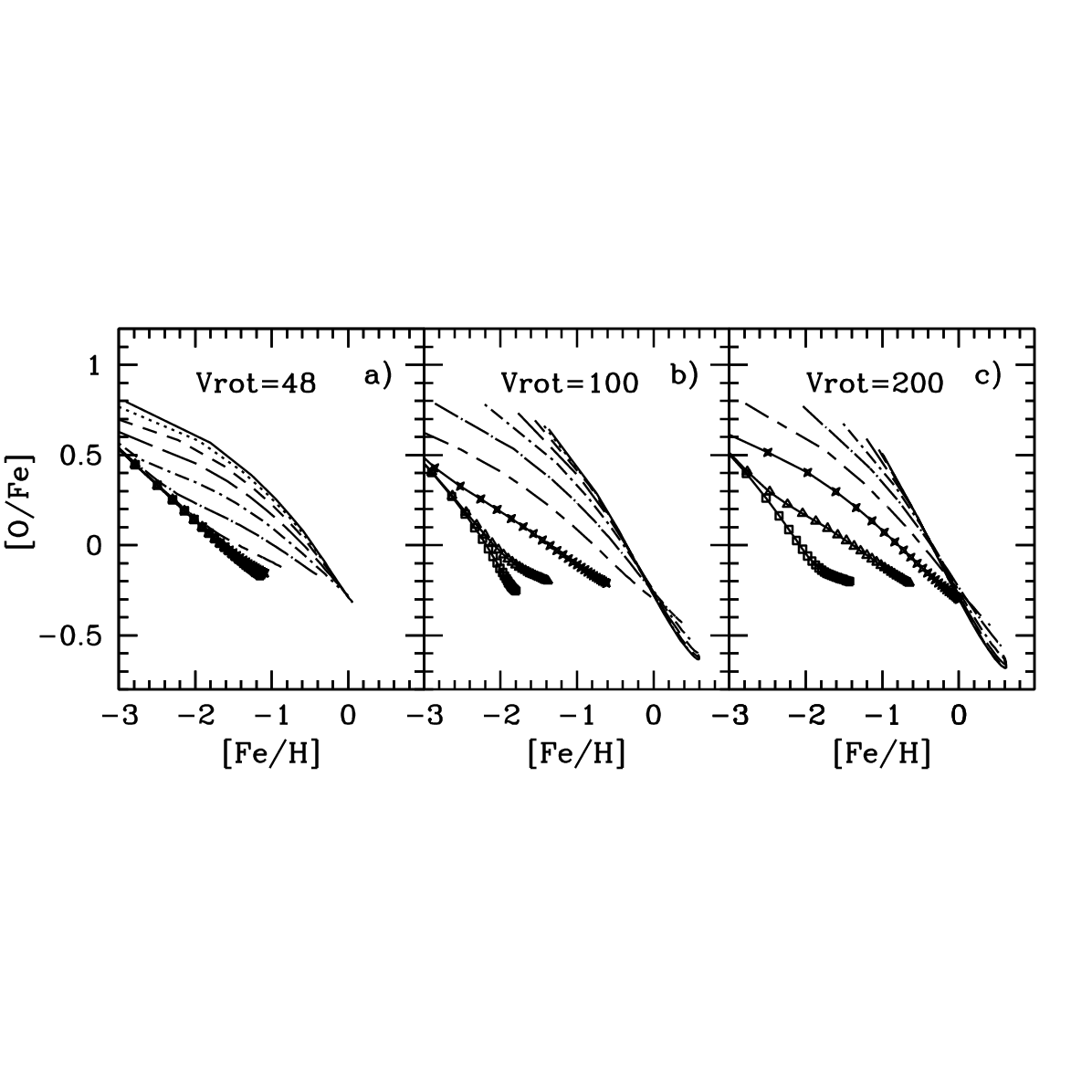}
\caption[]{Relative abundances [O/Fe] {\sl vs} [Fe/H]
for 3 different mass distributions in each
panel. Ten different morphological type are shown in each 
panel. Line types have the same meaning than in Fig.~\ref{hi}}
\label{ofe}
\end{figure*}

The relative abundances among elements give information about the
scales of evolution in galaxies or regions where the star formation
rates are different.  When the high star formation rates produce very
rapidly a large number of stars, the oxygen ejected mostly by the more
massive stars appears in the interstellar medium at the first epoch of
the evolution.  The iron ejected by the low mass stars appears, at
least, 1 Gyr later.  If the star formation has ceased to create stars,
due, for example, to the consumption of the gas, stars cannot 
incorporate this iron in their interiors.  Thus the stellar abundance
[O/Fe] would have over-solar values.  If the star formation produces
stars at a lower rate, more continuously, they would be created with a
similar abundance in oxygen and iron, and therefore the mean relative
abundance will be around solar.

Our models produce a large number of stars when the morphological
types are early or when the galaxies are massive, which implies higher
infall rates and more rapid evolution in their disks. We can see this
effect in Fig.~\ref{ofe} where we show in each panel the resulting
[O/Fe] {\sl vs} [Fe/H] for a given radial distribution of total mass
and for the 10 morphological types.  It is clear that all type of
massive galaxies produce over-solar abundances [O/Fe], with values
$\sim 0.5$ -- 0.8, for metallicities $\rm [Fe/H] < -0.7$ dex. For the
low mass galaxies this occurs only for the earliest morphological
types, while the later ones have smaller values of [O/Fe]. For these
latter galaxies, the [O/Fe] ratio starts to decrease toward solar
values at lower metallicities ($\sim 2.5$ dex) than for the more massive
galaxies, which maintain their iron underabundances until higher
metallicities ($\sim -1.5$ dex)

The usual {\sl plateau} in the relation [O/Fe] {\sl vs}
[Fe/H], also found in our previous models, is steeper in this new
work.  The only possible explanation resides in the different
nucleosynthesis yields for the massive stars used now in comparison
to those used before: \citet{woo95} yields produce more iron than
the old yields \citep{woo86}, and therefore the ratio [O/Fe]
decreases, although smoothly, from the beginning. When the SN-I's
explosions start the iron increases suddenly and the slope of the
relation steepens abruptly. The change of slope indicates the time
when this occurs.

\subsection{The star formation rate}

Radial distributions of star formation rate surface density show an
exponential shape in the outer disk, but a less clear one in the inner
regions, where some models show a distribution flatter than the
molecular gas, and even, decreasing toward low values of the star
formation rate in the center.  Although at first look the
SFR radial distributions seem to be different from those obtained by
Kennicutt from H$\alpha$ fluxes, the comparison of some particular
models with galaxy data results acceptable, as we will see in Section
4.2.

This effect is due to the star formation law assumed in the multiphase
model which has two modes to form stars: the spontaneous one, depending
on radius through the efficiency to form molecular clouds,
stronger in the inner than in the outer disk; the stimulated
one, resulting from the interactions of molecular clouds with
the existing massive stars. Thus, the star formation rate depends on
the molecular gas surface density and also on the massive stars one.

\begin{figure*}
\plotone{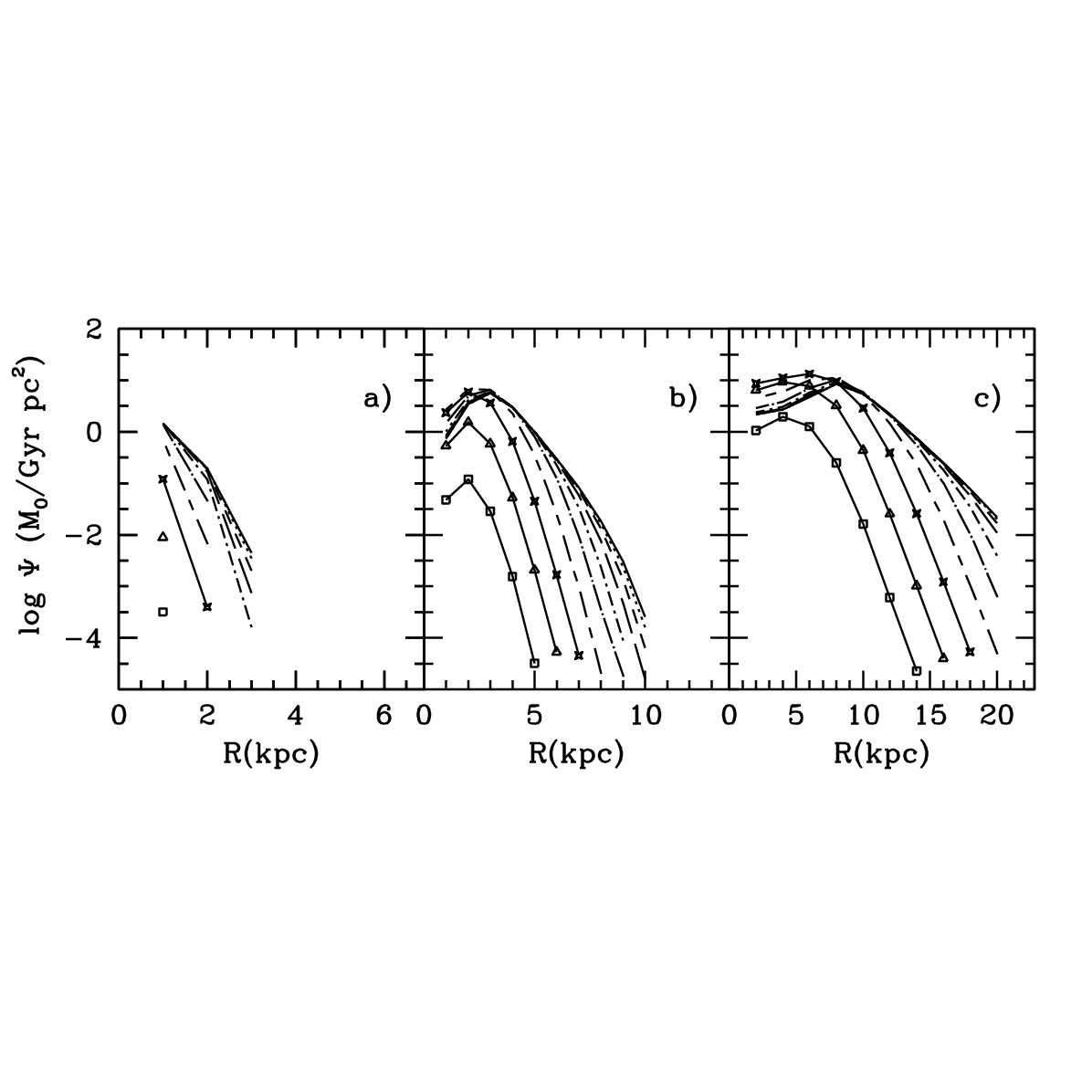}
\caption[]{Present epoch radial distributions of the surface density of
the star formation rate  for 3 different mass distributions 
Labels have the same meaning than in Fig.\ref{oh}}
\label{sfr}
\end{figure*}

\begin{figure*}
\plotone{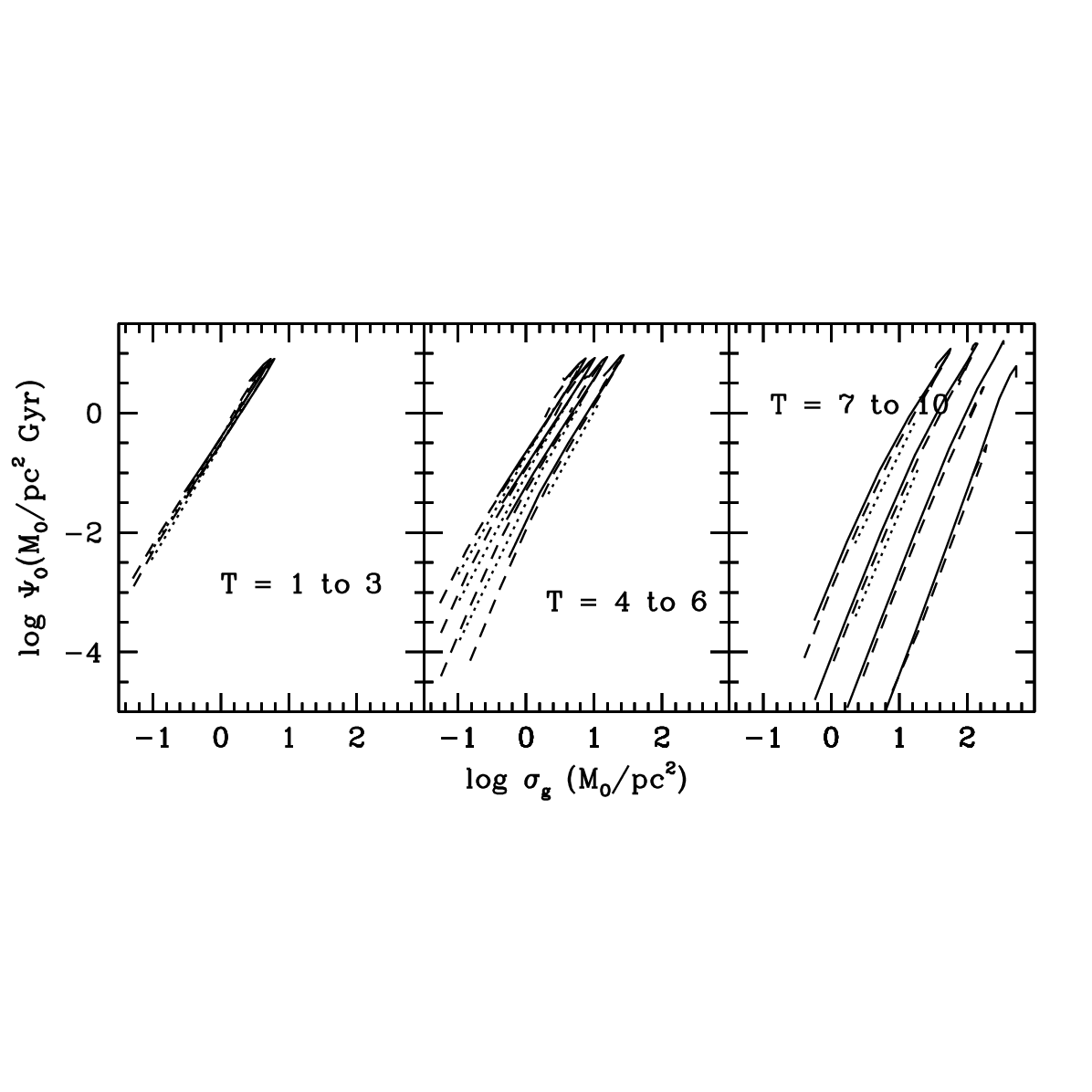}
\caption[]{The relation of the surface density of
the star formation rate  with the total gas density 
for 3 different mass distributions and different
morphological type as labeled}
\label{ken}
\end{figure*}

This relation reproduces the Kennicutt's relation \citep{ken89}, which
has two slopes when the SFR is represented {\sl vs} the total gas
surface density, Fig.~\ref{ken}. This means that a threshold in the
star formation is not strictly necessary since the same effect is
obtained when molecular gas and self-regulation are considered.

The star formation rate assumed in our models do not produce
bursts in the low mass galaxies of any morphological type.
 Only the massive  galaxies are able to  keep a large quantity
of gas in a small region, usually at the centers although sometimes
 at the inner disk regions. On the contrary, low mass
galaxies collapse very slowly, and thus the star formation rate maintains
a low level during the whole life of the galaxy. In fact  recent
works suggest the same scenario for both  low mass and  low surface
brightness galaxies \citep{dhoek00,leg00,bra01} in order to take into
account the observed data.  Our resulting abundances and gas fractions
are in agreement with these data, although the photometric
observations cannot still be compared with the results of our model.

\section{Analysis of model results: Tests}

With the selection of parameters and inputs described above, we have
ran a total of 440 models, with 44 different rotation curves
--implying 44 values of total mass, characteristic collapse time scale
and disk radius-- and 10 morphological types for each one of them,
implying 10 evolutionary rates for the star formation and gas
consumption in the disk.

For each model we obtain the time evolution of the halo and
the disk, and therefore the corresponding radial distributions for the
relevant quantities (masses, abundances, star formation rate, etc...).
The star formation history is followed for each radial region (halo
and disk, separately) and within each one, the mass in each phase of
matter: diffuse gas, molecular gas, low-mass stars, massive stars and
remnants, is also followed.  Besides that, we obtain abundances for 15
elements: H, D, He3, He4, C12, C13, N14, O16, Ne, Mg, Si, Ca, S, Fe
and neutron-rich nuclei in all radial regions for both halo and disk.

We now will compare the results with observational data in order to see
if models are adequately calibrated.

\subsection{Application to the MWG}

\begin{figure*}
\plotone{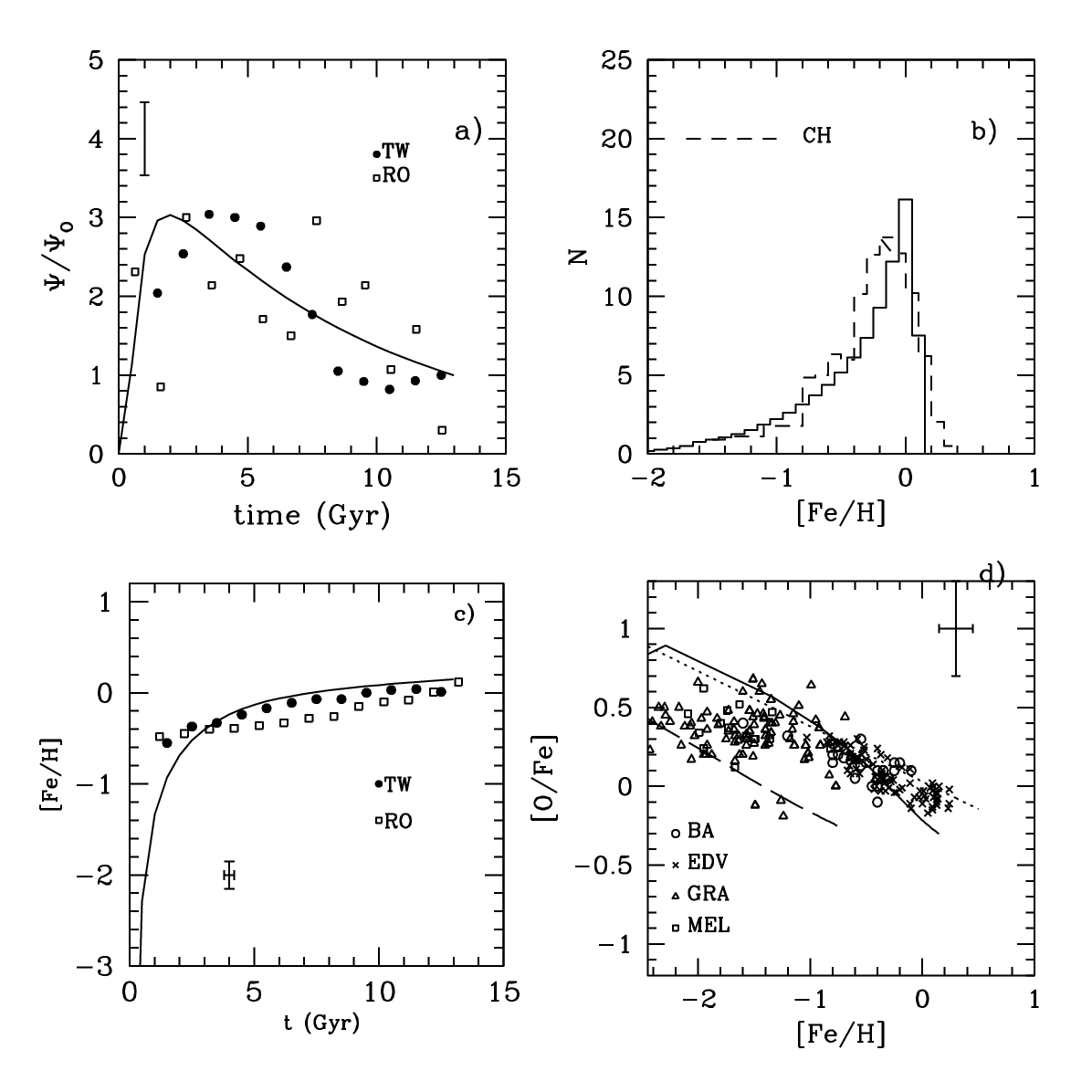}
\caption[]{The Solar Neighborhood disk evolution as results from the
chosen MWG model for the region located at $R=8$ kpc. a) The star
formation history with data from \citet{twa80} --filled dots:TW-- and
\citet{rocha00b} --open squares: RO--; b) The metallicity distribution
compared with the observed distribution from \citet{chang99}; c) The
age-metallicity relation with data from \citet{twa80} --filled
dots:TW-- and \citet{rocha00} --open squares: RO--; and d) the
relation [O/Fe] {\sl vs } [Fe/H] for the disk (solid line) and for the
halo (long dashed line) region models with data \citet{bar89} --open
dots: BA--, \citet{edv93} --crosses:EDV, \citet{grat00} --open
triangles: GRA--, and \citet{mel01} --open squares: MEL--. The dotted
line is the mean relation given by \citet{boe99}.}
\label{mwg1}
\end{figure*}

The first application of any theoretical model is to check its
validity to the MWG. A large set of observational data for the Solar
Neighborhood and the galactic disk exists and therefore the number of
constraints is large compared to the free parameters of the computed
models.

The model for $T=4$, and total mass distribution number 28,
$\lambda=1.0$ and maximum rotation velocity $V_{max}=200 km s^{-1}$, has
been chosen for comparison with the MWG.

The results corresponding to this model are represented in
Figs.~\ref{mwg1} and ~\ref{mwg2} together with the available data. In
Fig.~\ref{mwg1} we show the results for the disk Solar Neighborhood,
that is, the disk region located at R=8 kpc from the
Galactic center. The left panels a) and c) show the star formation
history and the age-metallicity relation. Both graphs reproduce
adequately the observed trends, although the model predicted star
formation maximum appears slightly shifted to earlier times with
respect to observations, and the high iron abundances observed at
times earlier than 1 Gyr are not fitted by the model.

Right panels b) and d) show the disk metallicity distribution and the
[O/Fe] {\sl vs} [Fe/H] relation. Both figures provide information
about the scale of star formation and enrichment. They also show a
reasonable agreement with available observational data. The metallicity
distribution resulting of the model does not show the G-dwarf problem: as
we can see, there is no excess of metal-poor stars. 

In the last panel d) data refer to disk (open dots and crosses)
and halo (triangle and squares) stars, while the solid line refers to
the disk region model which falls on the disk data locus.  The model
also predicts higher values of O/Fe for low metallicities but, taking into
account the stellar metallicity distribution, not many disk stars are expected
in this location in the diagram.

The data corresponding to the halo stars show two different behaviors
depending on how measurements are made.  We have included the mean
relation given by \citet{boe99}, --dotted line--, who gives a review
discussion about the subject and recalculates some of these
abundances, but these values are doubtful and under debate. The halo
model (long dashed line) shows a lower value, around 0.5 dex, as shown
by recent data from \citet{grat00} and \citet{mel01}.  For larger
metallicities this value decreases, but, as in the previous comparison
, there are no many halo stars in this locus, as the star formation in
the halo decreases very quickly after the first Gyr.

Our models do not cover the region where the highest metallicity
halo stars lie. This is probably due to the fact that the thick disk is
not considered as a separate component. In fact, there exists a continous
sequence of stellar populations along the vertical direction that would imply
a continous star formation and the existence of stars located
between our two model (halo and disk) lines.

\begin{figure*}
\plotone{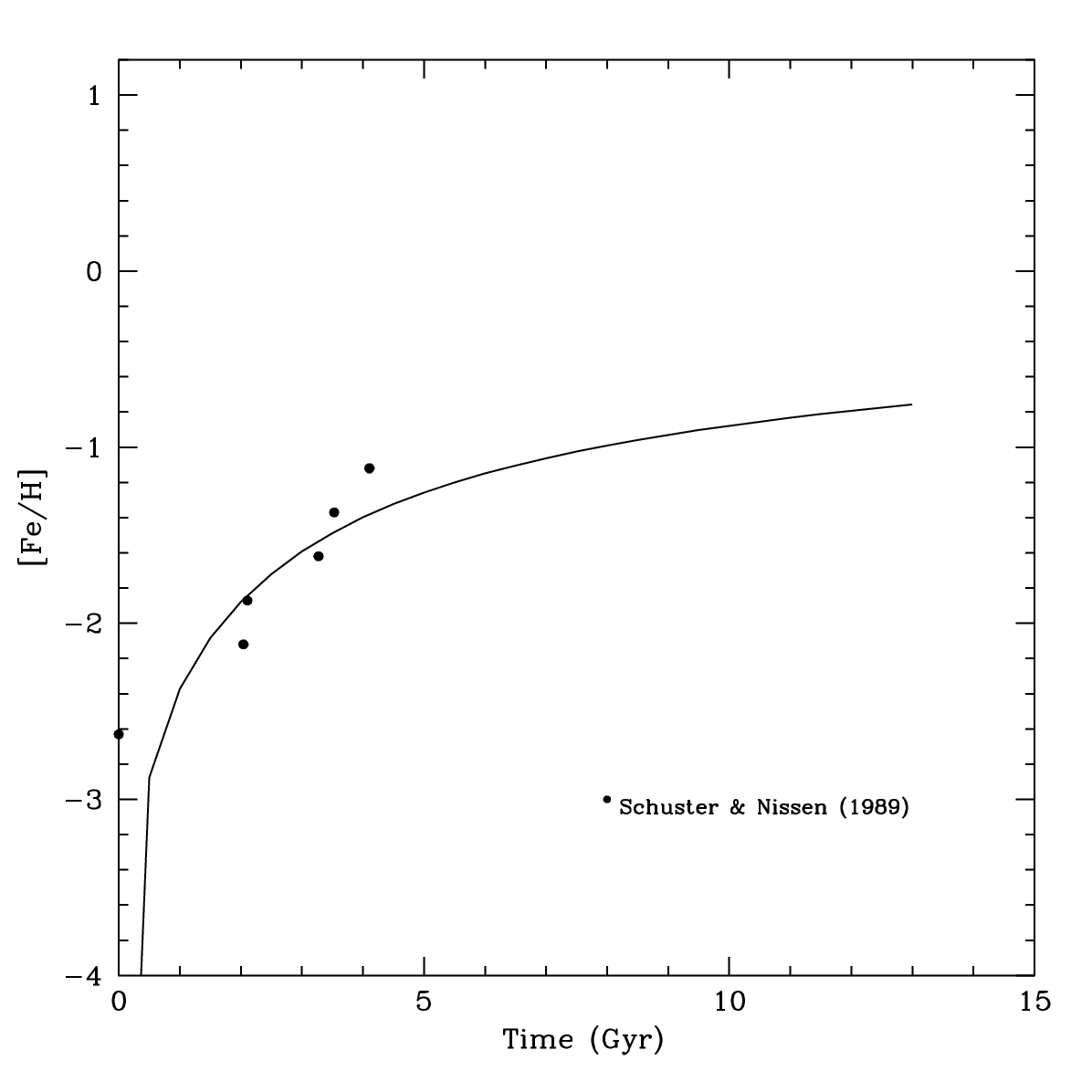}
\caption[]{The Solar Neighborhood halo region evolution 
as results from the chosen MWG model for the halo region 
located at $R=8$ kpc: The age-metallicity relation with 
data from \citet{schu89} --filled dots:SN.}
\label{halo}
\end{figure*}

\begin{figure*}
\plotone{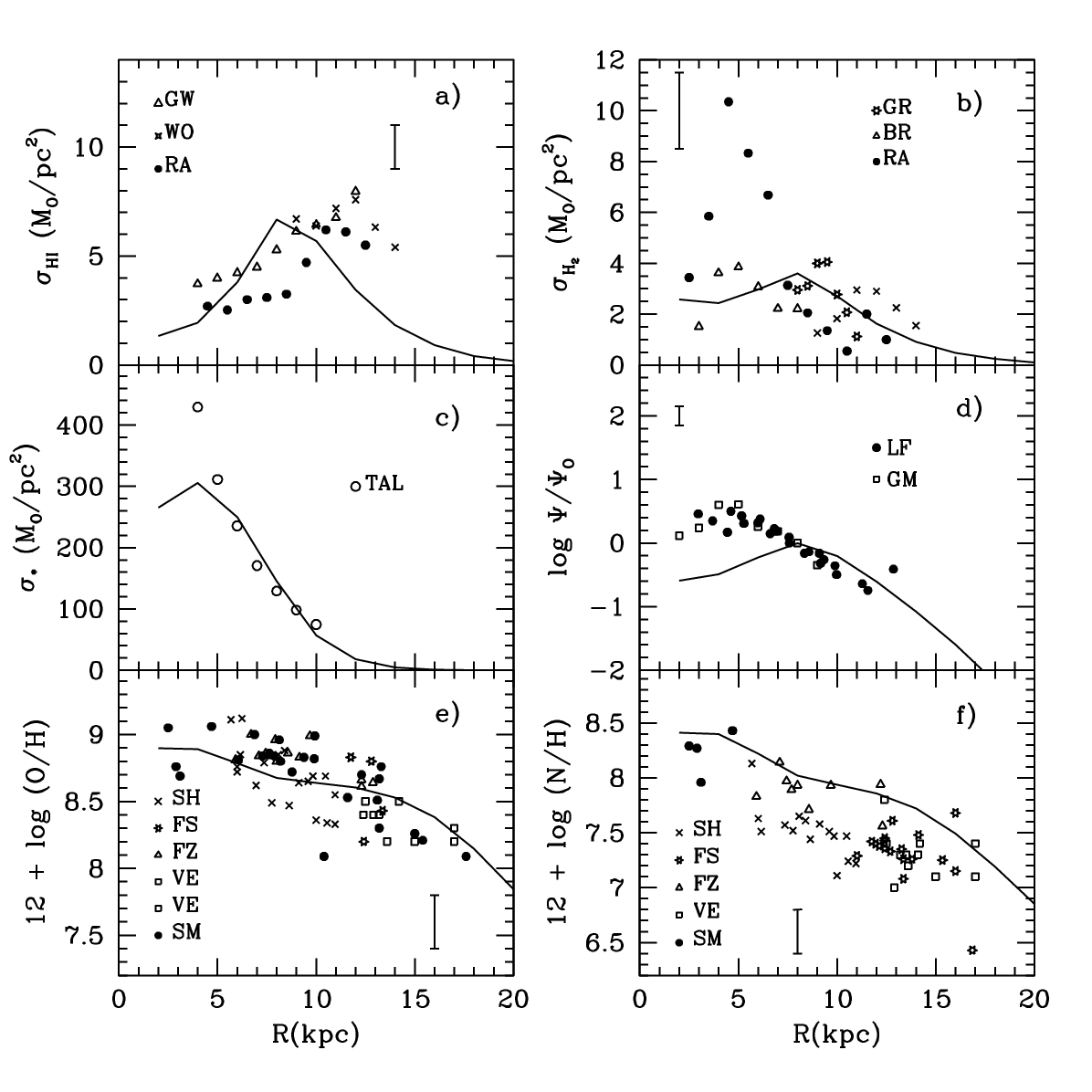}
\caption[]{Present epoch radial distributions for the MWG simulated
galaxy (distribution number 28,$\lambda= 1.00$, $T=4$): a) atomic gas
density with data from \citet{gar89} --open triangles: GW--,
\citet{wou90} --crosses: WO--, and \citet{rana91} --filled dots:RA--;
b) molecular gas surface density with data from \citet{grab87}
--stars: GR--, \citet{bron88} --open triangles:BR--, and
\citet{rana91} --filled dots: RA--; c) stellar surface density with
data from \citet{tal80}; d) star formation rate surface density
normalized to the present time solar value in logarithmic scale
\citep[data taken from][filled dots --LC-- and open squares --GM--,
respectively]{lac85,gus83}; e) and f) oxygen and nitrogen abundance as
$\rm 12+log (X/H)$ with data from \citet{sha83} --crosses:SH--,
\citet{fich91} --stars: FS--, \citet{fitz92} --open triangles:FZ--,
\citet{vil96}, --open squares: VE-- and \citet{sma97,sma01}
--stars:SM.}
\label{mwg2}
\end{figure*}

 Our treatment allows star formation in the galaxy barionic
halos which produces an certain level of enrichment. In the case of
our galaxy, this enrichment is in agreement with data for the Solar
Neighborhood halo. This is shown in Fig.~\ref{halo} where we plot the
evolution of the halo for the region located at R$=8$ kpc, for the
model representing the MWG. We can see that the age-metallicity
relation of the halo reproduces the data from \citet{schu89}.

Moreover, the ratio $\rm Mhalo/Mdisk$ must reproduce the
value given by observations \citep{san87} as we have already
explained in Section 2.  Since we only deal with two components
including the thick disk in the halo (or in the disk) region, the
fraction Mhalo/Mdisk must be around 0.10. In our model 28, T$=4$, for
R$=8$ kpc, the halo to disk stellar mass ratio is around 1/9 in
agreement with those estimates.

Fig.~\ref{mwg2} shows the radial distributions of diffuse and
molecular gas, stars, total mass, star formation rate and oxygen and
nitrogen abundances for the galactic disk.  The shape of the observed
radial distribution is well reproduced for the diffuse gas, with a
maximum in the outer disk.  For the total mass it is an exponential
distribution in agreement with the surface brightness distribution,
and radial gradients of abundances similar to those shown by the most
recent data.

For the molecular gas density is a quasi-exponential law, which
decreases at the inner disk regions (R$\le 7$kpc).  This decreasing
may be eliminated if we assume a decreasing of the cloud-cloud
collision process efficiency in those zones, but then the oxygen
abundance would also decrease in the center of the disk, in
disagreement with observations. On other hand it is necessary to
remind that the molecular masses are estimated from the CO intensity
through a calibration factor. This factor depends on metallicity
\citep{ver95,wil95} in a way which would produce smaller molecular gas
densities at the inner galactic disk if it is used. By taking into
account that recent data give low densities at these inner regions, we
consider that our model results may represent adequately the reality.

The radial distributions for oxygen and nitrogen (and also
for other element not shown here) show the usual variation with
galactocentric radius with a radial gradient in the range of what it
is observed. Although the fact of reproducing this characteristics
cannot be considered as a positive discriminant, this must be
nevertheless imperative for any one.  We show that our model
reproduces the data within the errors and dispersion range.  On the
other hand, this model shows a flattening in the inner disk which is
in agreement with the recent observations from \citet{sma01}.

The only feature which is not well reproduced is the SFR that
decreases toward the inner disk in apparent inconsistency with
observations. The star formation rate has a maximum in $R \sim 7-8$
kpc, while the observed star formation rate radial distribution
increases exponentially towards the center, or it has a maximum in 3-4
kpc.  In fact, it results difficult to explain why the star formation
rate is still so high at the inner disk, (in regions located at 3-4
kpc), where both phases of gas are already consumed.  The maximum of
the atomic gas density is around of 10-11 kpc, and the molecular gas
density has its maximum around 6 kpc. Negative feedback
caused by the increase of the cloud velocity dispersion migth be an
answer.  This effect is not taken into account in our models and may
delay the star formation in such way that the gas density remains high
for a longer time which, in turn, would imply a stronger star
formation rate at present. On the other hand,
recent data on the radial distribution of OB star formation in the
Galaxy \citep{bro00} shows a maximum around 5 kpc, decreasing towards
both sides. This may be an indication that the SFR indeed has a
maximum along the galactic disk, although the detection of young
sources, which are embedded in gas clouds, is difficult and might
result in a selection effect.  Nevertheless, a decrease in the star
formation radial distribution for the inner regions of disks is
observed in a large number of galaxies, as we will see in the next
section.

\subsection{Individual galaxies}

In what follows we analyze the radial distributions for gas, star
formation and abundances for some galaxies with large observational data sets
in order to check if the generic model reproduces the observed
characteristics of particular galaxies. 

\begin{deluxetable}{lcccccccccc}
\tablecaption{Galaxy  Sample Characteristics and Model Input Parameters.
\label{sample}}
\tablehead{
\colhead{Galaxy} & \colhead{T} & \colhead{Type}  & \colhead{D} 
& \colhead{Vel$_{\rm rot,max}$ }&
\colhead{R$_{0}$} &\colhead{$\tau_{0}$} & \colhead{$\epsilon_{\mu}$} &
\colhead{$\epsilon_{H}$} & \colhead{Mass Distr.}\\  
\colhead{Name}   &\colhead{}   & \colhead{Class} & 
\colhead{(Mpc)} &\colhead{($\rm km s^{-1}$)}& \colhead{(kpc)} & \colhead{(Gyr)}
 &\colhead{} & \colhead{} &  Number}
\startdata
NGC~300  & 7 & Scd/Sd  & 1.2  & 85  & 2.3 & 13.3 & 8.62E-02 & 2.18E-03 & 13 \\
NGC~598  & 6 & Sc/Scd  & 0.9  & 110 & 2.9 & 10.3 & 1.58E-02 & 1.11E-02 & 20\\
NGC~628  & 5 & Sc      & 11.4 & 220 & 7.7 & 3.28 & 2.86E-01 & 4.39E-02 & 32\\
NGC~4535 & 5 & ---     & 16.8 & 210 & 7.4 & 3.50 & 2.86E-01 & 4.39E-02 & 31\\ 
NGC~6946 & 6 & Scd     & 5.9  & 180 & 6.2 & 4.26 & 1.58E-02 & 1.11E-02 & 25\\ 
\enddata
\end{deluxetable}

The characteristics and corresponding input parameters of this
representative galaxy sample are given in Table~\ref{sample}. For each
galaxy, Column (1), the morphological type index is given in Column
(2), while the classical Hubble type is given in Column (3). The
adopted distance (taken from references following Table~\ref{data}, Column 2)
is in Column (4), the maximum rotation velocity is
given in Column (5) and the characteristic radius in Column (6). The last
columns, (7) to (9) give the collapse time scale and efficiencies for
molecular cloud and star formation. The last column is the number of
the radial distribution of total mass --corresponding to the column
(1) of Table~\ref{grid_m}-- chosen for each galaxy.

\begin{figure*}
\plotone{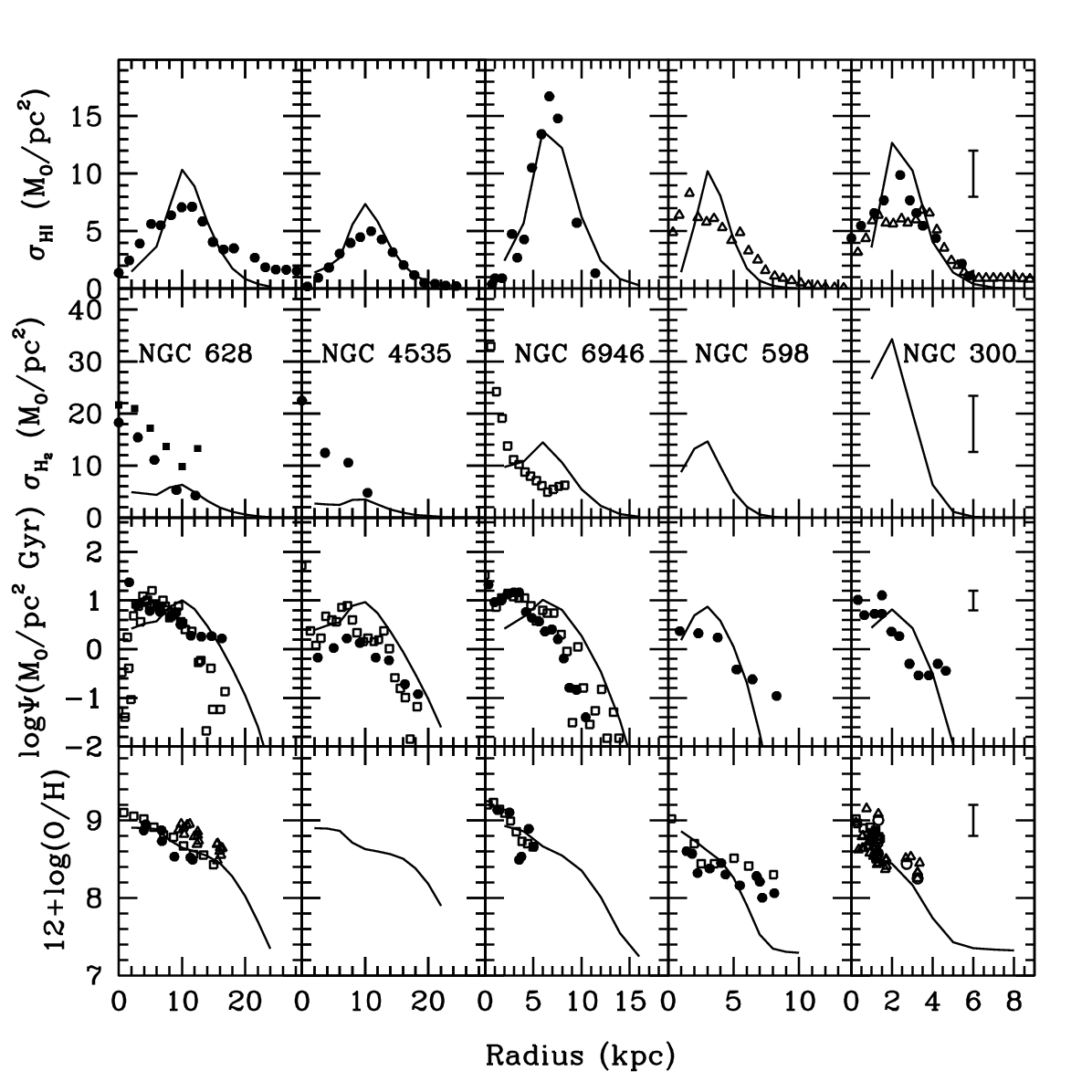}
\caption[]{Present epoch radial distributions for atomic and molecular
gas densities, (first and second rows), star formation rate (third
row), and oxygen abundance $\rm 12+log (O/H)$ (last row) for the
sample used to check the grid of chemical evolution models
(Table~\ref{sample}).  The observational data are taken from
references given in Table~\ref{data}.}
\label{chequeo}

\end{figure*}

The radial distributions of the different quantities for these
galaxies are shown in Fig.~\ref{chequeo} together with the
corresponding observational data, taken from references given in 
Table~\ref{data}.

In the first row of panels of Fig.~\ref{chequeo} we can clearly see
that the radial distributions of neutral hydrogen are well reproduced
by the models. We would like emphasize that, besides the fact that
there are no models which try to reproduce the radial distribution of
galaxies different than the ours, this comparison with diffuse gas
data has not been performed by any chemical evolution model because,
usually, the total gas, not the diffuse and molecular gas separately,
is used for fitting models. 

For each galaxy the distribution shows a maximum along the disk. For
galaxies with similar total mass this maximum is higher for later
morphological type, (NGC~4535 and NGC~6946).  For galaxies of the same
T, but different total mass, such as NGC~6946 and NGC~598, this
maximum does not change its absolute value, but it is located at a
different galactocentric distance, closer to the galaxy center for the
less massive one. This shows the effect of the collapse time scale,
which is longer for the less massive galaxies, thus producing a slower
evolution.

We would like clear that this good fitting of data, it is essentially
obtained when the most recent data and the adequate distance are used.
If old worst data are taken for the comparison with models, the fit
does not result good enough. In the same way the selection of the most
recently obtained distances improves extraordinarily the adjustment
between model and  observations. This same effect is noted for all the
other panels of this same Fig.~\ref{chequeo}

The radial distributions of molecular cloud surface density are shown
in the second row of Fig.~\ref{chequeo}. They show a shape similar to
an exponential function in the outer disk for all galaxies, but they
decrease at the inner ones. We might decrease the efficiencies
$\epsilon_{H}$ in these zones, and thus recover the exponential
shape. But in this case oxygen abundances are smaller than observed in
these same regions.  Differences between models and data are larger
than in the case of the diffuse gas, which is not unexpected, given
the larger uncertainties involved in the derivation of molecular
hydrogen masses. The non-dynamical treatment of the model
might be the cause of these differences. However, we stress that the
most recent data of molecular gas radial distributions \citep{nish01}
also show a decline of the surface density for the inner disks of
normal galaxies, (only barred galaxies also show, besides this hole, a
strong increase of the density in the center). Our models reproduce
this characteristic even if dynamical effects are not taken into
account. 

The radial distribution of the star formation rate for each galaxy is
 shown in the third row of Fig~\ref{chequeo}. We see that most of the
 distributions show a maximum near the center, but not always in the
 very central region. When the galaxy has evolved rapidly, the star
 formation rate has decreased in the inner disk, and the maximum has
 moved toward the external zones. If the galaxy is of late type or
 its mass is small, the evolution is slower and the star formation
 still shows a considerable level in the central regions of the disk,
 producing a quasi-exponential function for the star formation rate.
 The absolute values in the Maximo of the star formation rate
 histories are higher for the most massive galaxies.

In the last row of Fig.~\ref{chequeo}, we show the oxygen abundance
radial distribution for each galaxy with the corresponding
observational data.  The observed trend of steeper radial
distributions -- larger radial gradients-- for the late type galaxies
is reproduced by the models due to a slower collapse combined with
lower efficiencies for cloud and star formation rates, which defer the
creation of stars and the ejection of chemical elements to the
interstellar medium. On the other hand, the strength of the spiral arm
is taken into account by the radial dependence of parameter $\mu$
which is larger for the inner regions of the disks.  The star
formation rate results higher in the inner disk, thus producing a
radial gradient of abundances.

\begin{deluxetable}{lccccc}
\tablecaption{Galaxy Sample Data References.
\label{data}}
\tablehead{
\colhead{Galaxy}  & \colhead{D}  & \colhead{H\sc {i}} 
& \colhead{H$_{2}$}  & \colhead{SFR} & \colhead{[O/H]}}
\startdata
NGC~300  & 21 & 17,18 & \nodata & 7  & 3,7, 16\\ 
NGC~598  & 11 & 5     & \nodata & 10 & 12,23 \\
NGC~628  & 13 & 24 & 1,19 & 13,15 & 2,14,22 \\
NGC~4535 & 21 &  6 & 9 & 10,13 & \nodata \\
NGC~6946 & 8  &  4 & 17 & 10,13 & 2,14 \\
\enddata
\footnotesize{
(1)\citet{adl89}; (2)\citet{bel92}; (3)\citet{chris97}; 
(4)\citet{car90}; (5)\citet{cor00}; (6)\cite{cay90}; (7)\citet{deh88};
(8)\citet{kar00}; (9)\citet{kenn89}; (10)\citet{ken89}; 
(11)\citet{kim02}; (12)\citet{kwi81}; (13)\citet{mar01}; 
(14)\citet{mcc85};(15)\citet{nat92}; (16)\citet{pag79}; (17)\citet{puc90};
(18)\citet{rog79}; (19)\citet{sage89}; (20)\citet{tac86};
(21)\citet{tul88}; (22)\citet{vzee98}; (23)\citet{vil88};
(24)\citet{wev86}}
\end{deluxetable}

\section{Conclusions}

We have run 420 models within the multiphase framework, corresponding
the combination of to 42 different mass radial distributions and 10
evolutionary rates, corresponding to 10 efficiency values for the
molecular cloud and spontaneous star formation processes,
respectively.

The results obtained in this bi-parametric grid may be compared with
any galaxy of given total mass (or equivalently, maximum rotation
velocity) and morphological type.  These results include for each
radial region of halo and disk, the total mass included in the region,
the star formation rate, the mass in each phase, diffuse and molecular
gas, low and high mass stars, remnants, (types I and II supernova
rates, and abundances for 15 elements.

The models reproduce adequately most of the global characteristics
observed in spiral galaxies along the Hubble Sequence.  In
fact, taking into account that, with the exception of the calibration
done with the Milky Way galaxy (MWG), the grid has not been computed
to compare with specific galaxies, the perfomed comparison between
model results with data must be considered very good.
Actually, to our knowledge, there are not other chemical evolution
models which compare predicted with observed radial distributions of
diffuse and molecular gas, star formation rate and abundances for
disk galaxies other than MWG.  In this sense our models should be
considered as an improvement over the standard ones.

\begin{enumerate}

\item The atomic gas shows a maximum in its radial distribution for all
galaxies. This maximum is nearer to the center in the late and less
evolved galaxies than in the more massive galaxies, for which the maximum
is along the disk and moving toward the outer zones. This behavior produces,
in some cases, a {\sl hole} in the central zone for the diffuse gas

\item The molecular gas evolves with a time delay with respect to that
 of the atomic gas. Thus, the maximum of this distribution lasts longer in
the center than in the case of the diffuse gas. A central maximum
 and an exponential function for the molecular gas  are found as
 usually  observed.

\item The $\rm H_{2}/M_{gas}$  ratio  increases from the late to early type
galaxies  as the efficiency of the formation of
molecular clouds also increases. However, models seem to indicate that
the earliest and most massive galaxies have already consumed some of
this molecular gas, showing a decrease in the ratio on the
disk, in comparison with intermediate type galaxies.

\item The oxygen abundance  reaches a maximum level, as a consequence of an
saturation effect which  occurs earlier for the massive and early
type galaxies. The less evolved galaxies  do not reach
this saturation level, except in  the central region, and therefore show
a steep radial gradient. This correlation is in
agreement with  data \citep{vil92,zar94,gar97}. The effect of this
saturation in the oxygen abundance has also been recently observed
in the inner regions of the Galaxy \citep{sma01}.

\item The less massive and latest type galaxies have not yet developed
a radial gradient of oxygen abundances, and show flat radial
distributions. This simulates an effect on-off: for $T=7$ a radial
gradient appears if $\lambda > 0.15$ while at $T=8 $ it only appears
for $\lambda \sim 1.50$.  This behavior is in agreement with
observations and solves the apparent inconsistency shown by trends
giving steep gradients for late type galaxies and flatter ones for the
earliest ones, while, at the same time most irregulars show no
gradient at all and very uniform abundances.

\end{enumerate}

The time evolution of these models as well as 
the spectrophotometric evolution are obvious steps forward in this kind of
works and will be carry out in the near future.

\begin{acknowledgements}

We thank the anonymous referee for the large number of useful suggestions.
This work has been partially funded by the Spanish Ministerio de
Ciencia y Tecnolog\'{\i}a through the project AYA-2000-093, 
Italian MURST and NATO Collaborative Linkage Grant PST.CLG.976036. This work
has made use of the Nasa Astrophysics Data System, and the NASA/IPAC
Extragalactic Database (NED), which is operated by the Jet Propulsion
Laboratory, Caltech, under contract with the National Aeronautics and
Space Administration.

\end{acknowledgements}

\end{document}